\documentclass[
reprint,
superscriptaddress,
 amsmath,
 amssymb,
aps,
]{revtex4-2}

\usepackage{natbib}
\usepackage{graphicx}
\usepackage{dcolumn}
\usepackage{bm}
\usepackage{colortbl}
\usepackage{color,soul}
\usepackage{gensymb} 

\begin{document}


\title{Magnetic field control of light-induced spin accumulation in monolayer MoSe$_2$}

\author{Rafael R. Rojas-Lopez}
  \email{rrlopez@fisica.ufmg.br}
\affiliation{Departamento de Física, Universidade Federal de Minas Gerais, 31270-901, Belo Horizonte, Brazil.}
\affiliation{Zernike Institute for Advanced Materials, University of Groningen, 9747 AG Groningen, The Netherlands.}
\author{Freddie Hendriks}%
\affiliation{Zernike Institute for Advanced Materials, University of Groningen, 9747 AG Groningen, The Netherlands.}
\author{Caspar H. van der Wal}%
\affiliation{Zernike Institute for Advanced Materials, University of Groningen, 9747 AG Groningen, The Netherlands.}
\author{Paulo S. S. Guimarães}
\affiliation{Departamento de Física, Universidade Federal de Minas Gerais, 31270-901, Belo Horizonte, Brazil.}
\author{Marcos H. D. Guimarães}%
   \email{m.h.guimaraes@rug.nl}
\affiliation{Zernike Institute for Advanced Materials, University of Groningen, 9747 AG Groningen, The Netherlands.}

\begin{abstract}
Semiconductor transition metal dichalcogenides (TMDs) have equivalent dynamics for their two spin/valley species. 
This arises from their energy-degenerated spin states, connected via time-reversal symmetry. 
When an out-of-plane magnetic field is applied, time-reversal symmetry is broken and the energies of the spin-polarized bands shift, resulting in different bandgaps and dynamics in the K$_+$ and K$_-$ valleys. 
Here, we use time-resolved Kerr rotation to study the magnetic field dependence of the spin dynamics in monolayer MoSe$_2$. 
We show that the magnetic field can control the light-induced spin accumulation of the two valley states, with a small effect on the recombination lifetimes.  
We unveil that the magnetic field-dependent spin accumulation is in agreement with hole spin dynamics at the longer timescales, indicating that the electron spins have faster relaxation rates. 
We propose a rate equation model that suggests that lifting the energy-degeneracy of the valleys induces an ultrafast spin-flip toward the stabilization of the valley with the higher valence band energy. 
Our results provide an experimental insight into the ultrafast charge and spin dynamics in TMDs and a way to control it, which will be useful for the development of new spintronic and valleytronic applications. 
\end{abstract}

\maketitle                


Atomically-thin transition metal dichalcogenides (TMDs) offer the possibility to optically address the spin and valley degrees-of-freedom of charge carriers.
Through circularly polarized light, $\sigma_{+}$ or $\sigma_{-}$, one can excite electron-hole pairs at opposite points at the edges of the Brillouin zone, known as $K_{+}$ and $K_{-}$ valleys \cite{Mak2012,Zeng2012,Xu2014}.
The high spin-orbit coupling in these materials further causes a spin splitting of the levels, connecting the spin and valley degrees-of-freedom and their dynamics.
For this reason, TMDs are very attractive for (opto)valleytronic and (opto)spintronic applications \cite{Mak2016, Liu2016, Schaibley2016, Zhong2017, Luo2017}.
One of the main bottlenecks for such applications is the control of these degrees-of-freedom.
Magnetic fields have been shown to strongly affect the valley polarization.
This has been demonstrated by photoluminescence (PL) measurements of TMD monolayers under high out-of-plane magnetic fields, showing a linear shift of the emission peaks reflecting the reduction of the band gap for one valley and the increase of the other \cite{Macneill2015, Li2014, Srivastava2015, Aivazian2015, Koperski2019}.
This effect became known as the valley-Zeeman effect.
These experiments extract an exciton and trion g-factor of $\sim$ $-$4, which is in agreement with theoretical expectations \cite{Wang20152D, Rybkovskiy2017, Wozniak2020}.
Different works have reported the dynamics of electrons, excitons, and trions by performing time-resolved photoluminescence (TRPL) \cite{Wang2015, Godde2016, Wang2020}, time-resolved differential reflectivity (TRDR) \cite{Kumar2014, Kumar2014prb, Ye2018}, and time-resolved Kerr rotation (TRKR) \cite{Zhu2014, Hsu2015, Guimaraes2018, Anghel2018, Ersfeld2019, Li2020}, with most works focusing on either zero or in-plane external magnetic fields.
More recently, Zhang et al. \cite{Zhang2021} showed the effect of out-of-plane magnetic fields on polarized TRPL, but without any clear difference between the results obtained at different circularly polarized excitation and high magnetic fields.
Nonetheless, PL is solely sensitive to the valley polarization and radiative decay of excitons and trions.
Therefore, PL is insensitive to light-induced spin polarization of resident carriers which can sustain the spin information over much longer times. 
The understanding of the decay processes and possibilities of control of resident carriers are of huge importance for the engineering of a new generation of opto-spintronic devices \cite{Seyler2018, Benitez2018, Luo2017, Sierra2021}.
However, the use of an out-of-plane magnetic field to control the long-lived spin dynamics in monolayer TMDs -- beyond the radiative recombination times -- remains largely unexplored.

\begin{figure*}[ht!]
    \centering
    \includegraphics[scale=0.92]{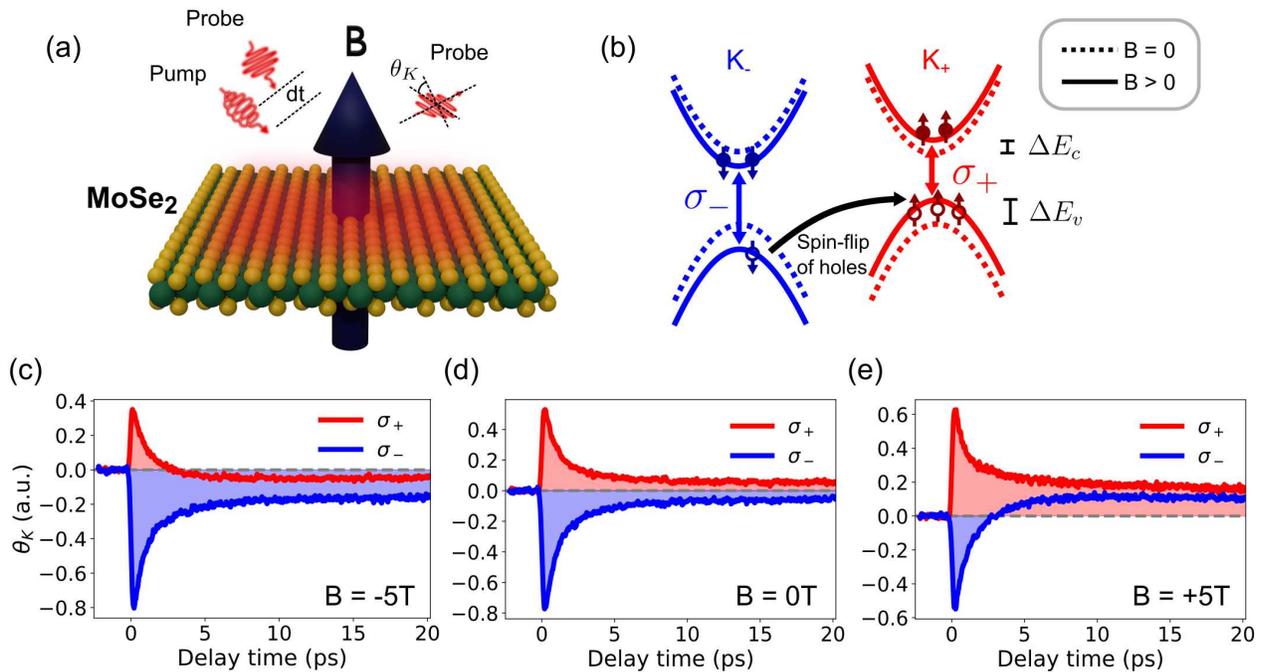}
    \caption{(a) Schematics of the TRKR measurements with an external magnetic field. (b) Valley-Zeeman effect on the K$_{+}$ and K$_{-}$ spin states of monolayer MoSe$_2$. Dashed lines indicates the zero magnetic field spin states while solid lines represent the magnetic field shifted states. The Valley-Zeeman effect results in larger energy shifting in the valence band $\Delta E_v$ than in the conduction band $\Delta E_c$. (c) TRKR signal for $\sigma_+$ (red) and $\sigma_-$ (blue) polarized pump, at different magnetic fields: B = -5 T, 0 T (d), +5 T (e), for $\lambda$= 755 nm and T = 6 K.}
    \label{Fig1:summary}
\end{figure*}

Here we show that the spin information of long-lived carriers in monolayer MoSe$_2$ can be effectively controlled via the valley-Zeeman effect.
Using time-resolved magneto-optic Kerr effect, we show that magnetic fields can induce an ultrafast spin-valley scattering and enhancement of a light-induced spin accumulation with a preference for the valley with higher valence band energy.
While we find that the spin scattering rates do not show a clear trend for the external magnetic field strength, the magneto-optical signal strength shows a clear linear behavior.
Particularly, we observe a nearly field-independent long spin lifetime of $\sim$2 ns.
The magnetic field dependence of the light-induced spin accumulation we measure agrees with hole spin dynamics and resident carrier recombination at longer timescales, while we argue that electron spins may have faster relaxation rates.
Our results are well described by a simple rate-equation model, which takes into consideration a field-dependent carrier recombination and intervalley scattering.
Our model indicates that the magnetic field has a stronger effect on the hole transfer between valleys than in electrons, resulting in a spin imbalance with a larger population of carriers from the valley with the smaller bandgap. 
In this way, we demonstrate here that an applied magnetic field can be used to effectively control the optically-generated spin accumulation in TMDs.


Our samples were prepared by mechanical exfoliation of a bulk MoSe$_2$ crystal (HQ Graphene) onto a polydimethylsiloxane (PDMS - Gel Pack) and then transferred to a 285 nm SiO$_2$/Si substrate.
The spin dynamics in our samples was measured by TRKR, which presents a higher sensitivity to spin-related phenomena, compared with TRPL or TRDR.
We perform a single-color (degenerated) pump-probe technique in a dual-frequency modulation (see supplementary information) using a similar experimental setup as described elsewhere \cite{Guimaraes2018}.
When a high intensity circularly polarized ($\sigma _{\pm}$) laser pulse (pump) excites the monolayer, electrons of a single valley ($K_{\pm}$) are excited to the conduction band and can generate a spin imbalance.
By shining a linearly polarized (probe) pulse at a certain time-delay after the pump pulse, the spin/valley imbalance is measured by a change on the polarization axis of the reflected beam, which is rotated by an angle $\theta_{K}$ (Fig. \ref{Fig1:summary}a).
Finally, the time evolution of the spin imbalance in the sample is measured by changing the delay time (d$t$) between the pump and probe pulses.
All measurements were performed at low-temperature T = 6 K.


Figure \ref{Fig1:summary}d shows the TRKR at zero magnetic field and how the sign of the Kerr rotation depends on which valley we are exciting: a positive (negative) Kerr rotation is observed upon shining $\sigma _+$ ($\sigma _-$) polarized pump pulses.
This behavior is a fingerprint of the symmetry and spin selectivity of the two valleys of the TMD, when excited with circularly polarized light \cite{Zhu2014, Hsu2015, DalConte2015, Guimaraes2018, Li2020}.
Two decay time constants are visible in our measurements, a faster ($\tau_1 \sim$ 1.5 ps) and a longer one ($\tau_2 \sim$ 2 ns) that goes beyond the measurement range.

We observe a drastic change on the TRKR dynamics upon an applied magnetic field.
Figure \ref{Fig1:summary}c and e show the TRKR signal of our sample for an applied magnetic field of -5 T and +5 T, respectively.
Within the first 3 ps after excitation we observe a reversal of the TRKR signal for one pump polarization, while a stabilization for the other.
This can be explained by an out-of-plane magnetic field lifting the valley degeneracy causing an opposite shift on the TMD bands of opposite valleys \cite{Macneill2015, Li2014, Srivastava2015, Aivazian2015} (Fig \ref{Fig1:summary}b).
Due to the different net angular momenta of the valence and conduction bands, the energy shifts for the conduction ($\Delta E_c$) and valence bands ($\Delta E_v$) are also different.
The resulting effect is that the lowest energy state for holes lies within one valley, while for electrons lies in the other.

These observations reveal that our TRKR signal is dominated by hole relaxation.
Upon populating the $K_{-}$ valley through a $\sigma_{-}$-polarized pump, we observe a reversal of the signal from negative to positive for positive magnetic fields.
For this case, the electron ground state lies in the conduction band of the $K_{-}$ valley, while the ground state of holes is located at the $K_{+}$ valley.
The fact that we observe a positive TRKR signal after $\sim$ 3 ps, indicates a higher spin population of the $K_{+}$ valley, implying a fast intervalley scattering and that the spin accumulation measured in our experiments arises from the hole spin.
When the direction of the magnetic field is reversed, similar behavior is observed but with an opposite TRKR signal.

\begin{figure}[t]
    \centering
    \includegraphics[width=\linewidth]{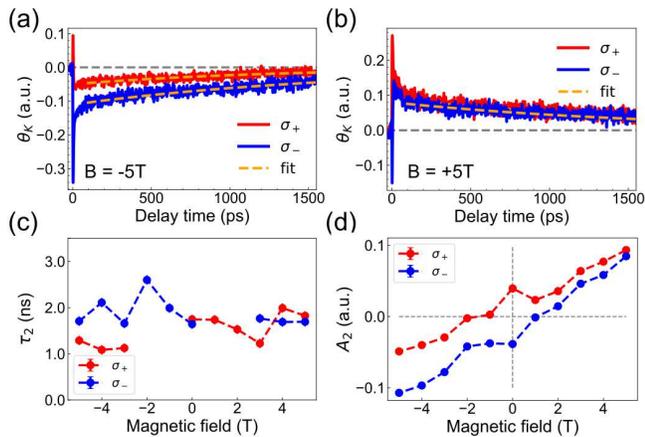}
    \caption{(a) TRKR signal for $\sigma_+$ (red) and $\sigma_-$ (blue) polarized pump, at B = -5 T (a) and +5 T (b) at long time scales. The orange dashed lines show the fit that extracts the slow relaxation times. (c) Extracted slow decay times ($\tau _2$) and (d) amplitudes (A$_2$) at different magnetic fields for the two excitation polarizations.}
    \label{Fig3:lifetimes}
\end{figure}

The long decay times we observe do not show any clear dependence with the magnetic field.
The two decay time constants are obtained through a biexponential decay of the form $\theta _K = A_1 e^{-t/\tau _1}+A_2 e^{-t/\tau _2}$, where $A_n$ are the Kerr rotation angles at a delay time d$t$ = 0 and $\tau _n$ the decay times at the fast and slow decay processes $n = 1, 2$, respectively.
This is done for measurements using both directions of pump polarization and various values of applied magnetic fields, from B = -5 T to +5 T.
Figures \ref{Fig3:lifetimes}a and \ref{Fig3:lifetimes}b show the TRKR signals at B = -5 T to +5 T over a long d$t$ range, up to 1.55 ns.
The TRKR signals are still clearly measurable even at these long-time delays.
The magnetic field dependence for $\tau_2$ is shown in Figure \ref{Fig3:lifetimes}c (see the supplementary material for $\tau_1$).
For zero, and also high magnetic fields, we find non-zero pump-induced TRKR signals.
However, for 0$< |B| <$ 3 T, the signal for one of the pump polarizations is within our experimental noise, and therefore only measurements for one pump polarization were used within this range.
Nonetheless, we do not observe any striking features on the TRKR decay times for the whole range of magnetic fields studied, even though additional spin scattering channels are predicted to arise from the breaking of time-reversal symmetry \cite{Gilardoni2021}.
This indicates that, while the intervalley scattering rate could be modified by the magnetic field, the main source of spin scattering remains unaffected for our samples.

The amplitude of our TRKR signals shows a clear linear behavior with the magnetic field, Fig. \ref{Fig3:lifetimes}d.
Strikingly, the slopes for the two pump polarizations are slightly different, resulting in a non-symmetric response with the magnetic field direction at long time delays.
Previous measurement runs of the same sample also presented a linear trend with the magnetic field, with different slopes for each excitation polarization, see supplementary material for details.
Although small experimental artifacts are not discarded, a phenomenological origin of these results would be inconsistent with a simple description involving solely a Zeeman energy shift of the energy states.
We currently do not have a clear understanding of the origin of such an effect, which should be explored in more detail in later studies.

\begin{figure}[t]
    \centering
    \includegraphics[width=\linewidth]{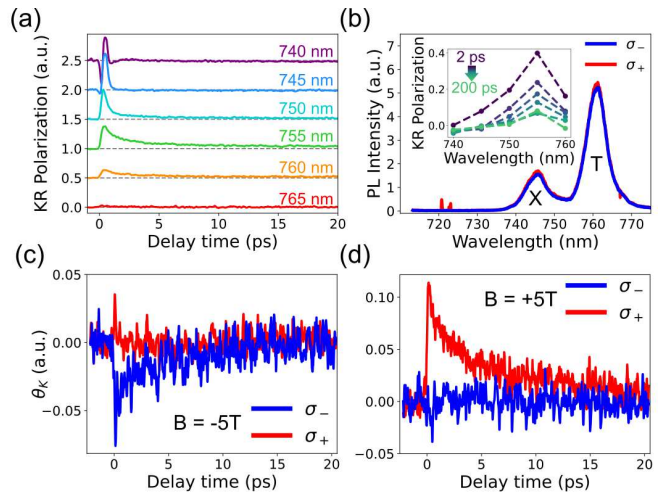}
    \caption{(a) TRKR Polarization at different wavelengths and zero external magnetic fields. Measurements were shifted in the vertical axis for clarity. (b) Polarized photoluminescence spectra of our sample at B = 0 T. Inset: Intensity profile of the TRKR Polarization at different delay times: 2, 4, 6, 10, 100, and 200 ps. (c) TRKR for $\sigma_+$ (red) and $\sigma_-$ (blue) polarized pump at $\lambda$ = 765 nm for B = -5 T (c) and +5 T (d) at T = 6 K. }
    \label{Fig2:wavelength}
\end{figure}

\begin{figure*}[t]
    \centering
	\includegraphics[scale=0.95]{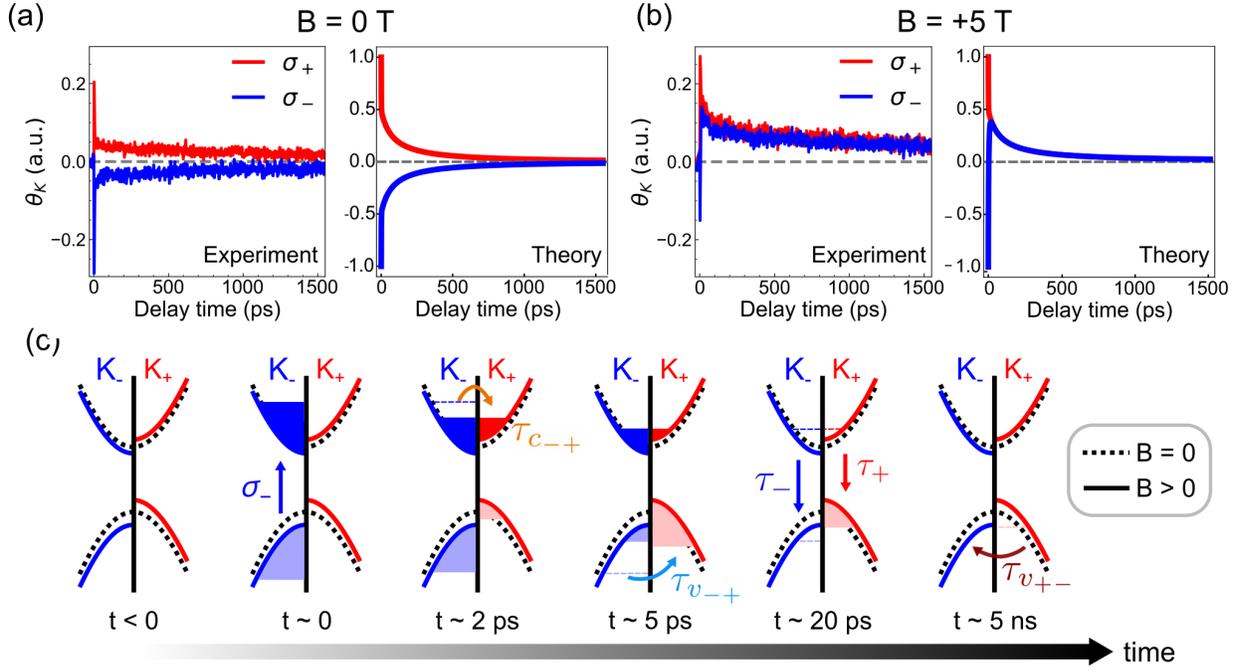}
	\caption{(a) Experimental (left) and theoretical (right) results for B = 0 T with $\tau_+=\tau_- =$ 20 ps, $\tau_{c_{+-}}=\tau_{c_{-+}} =$ 1.8 ps and $\tau_{v_{+-}}=\tau_{v_{-+}} =$ 5 ns. (b) Experimental (left) and theoretical (right) results for B = +5 T with an asymmetric hole scattering time $\tau_{v_{+-}} = 5$ ps, $\tau_{v_{-+}} =$ 5 ns. (c) Representation of the dynamics of the spin population under positive magnetic fields: the ground state of the system with an initial hole population is photo-excited ($\sigma _-$) at the valley $K_-$. Later, electrons are scattered between valleys in the conduction band followed by the scattering of holes from $K_-$ to $K_+$. Subsequently, excited states recombine radiatively. Finally, resident holes in $K_+$ reach thermal equilibrium with $K_-$. The represented transfer times indicate the dominant process at each panel.}
    \label{Fig4:model}
\end{figure*}

By studying the dependence of the TRKR signal on the excitation wavelength, we can probe the contribution of excitons and trions to the spin signal.
To reduce spurious effects which are independent of the valley/spin polarization, here we use the polarization of the Kerr rotation, namely the difference of the TRKR measurements for the two polarizations of excitation $(\theta_K(\sigma_+)-\theta_K(\sigma_-))/2$ \cite{Guimaraes2018}.
As expected, our signal is strongly modulated upon a change of the wavelength when going over the exciton and trion resonances (Fig. \ref{Fig2:wavelength}a).
We observe that when exciting with $\lambda \leqslant$ 745 nm, KR polarization decays in the first couple of picoseconds.
On the other hand, when exciting at $\lambda \geqslant$ 750 nm, the spin imbalance remains finite, within the time range of our measurements. 
The largest signal is seen at $\lambda=$ 755 nm, see inset of Figure \ref{Fig2:wavelength}b.
A comparison to photoluminescence measurements (Fig. \ref{Fig2:wavelength}b) reveals that our findings are consistent with short-lived signals coming from excitations at and below the wavelength of the exciton resonance (X), while the long-lived signals come from energies close to the peak of trion emission (T).
Our observations are in agreement with previous reports in literature \cite{Schwemmer2017, Godde2016, Hsu2015, Wang2015} showing that trions dominate the spin signal in TMDs monolayers.

We observe an interesting effect arising from the magnetic field-induced reduction of the bandgap for one valley with respect to the other.
For high magnetic fields (B = $\pm$5 T) and with the lowest excitation energy resulting on a measurable signal ($\lambda$ = 765 nm), we only observe a TRKR signal above our noise level for one pump polarization, Figure \ref{Fig2:wavelength}c and d.
For the positive magnetic field, the bandgap of the K$_+$ valley is reduced and produces a spin accumulation when excited with $\sigma_+$.
Nevertheless, as the bandgap at K$_-$ is increased, the energy of the pumping photons is not enough to excite the electrons and generate any spin imbalance, resulting in no TRKR signal for $\sigma_-$ excitation.
A similar behavior is observed for negative magnetic fields, but with opposite spin accumulation in the K$_-$ valley and no signal for $\sigma_+$ excitation.
Additional measurements of wavelength dependence with the magnetic field can be found in the supplementary material.

To quantitatively describe the spin dynamics and to elucidate the relaxation processes behind our measurements, we use a rate-equation model similar to what has been proposed before \cite{Hsu2015}, but including the effect of an external magnetic field.
We consider three main carrier scattering processes: direct radiative recombination at the same valley, the spin-valley flip of electrons in the conduction band, and the spin-valley flip of holes in the valence band.
The breaking of the energy degeneracy of the $K_+$ and $K_-$ valleys is represented by different intervalley scattering rates.
Therefore, considering the more general case, where conduction and valence band transfer rates, and also radiative recombination of both valleys, are different, we get to the set of equations

\begin{eqnarray}
    \frac{dn_+}{dt}&=&-\frac{n_+p_+}{\tau_+}-\frac{n_+}{\tau_{c_{+-}}}+\frac{n_-}{\tau_{c_{-+}}} \\
    \frac{dn_-}{dt}&=&-\frac{n_-p_-}{\tau_-}-\frac{n_-}{\tau_{c_{-+}}}+\frac{n_+}{\tau_{c_{+-}}} \\
    \frac{dp_+}{dt}&=&-\frac{n_+p_+}{\tau_+}-\frac{p_+}{\tau_{v_{+-}}}+\frac{p_-}{\tau_{v_{-+}}} \\
    \frac{dp_-}{dt}&=&-\frac{n_-p_-}{\tau_-}-\frac{p_-}{\tau_{v_{-+}}}+\frac{p_+}{\tau_{v_{+-}}} 
\end{eqnarray}

\noindent where $n_i$ and $p_i$ are the photo-excited populations at the $K_i$ valley for electrons and holes, respectively.
The first term at the right side of each equation represents the radiative recombination at a specific valley with recombination time $\tau _\pm$.
The second term represents the population of carriers that is transferred to the opposite valley in the conduction or valence bands with an intervalley scattering time $\tau _{c,v_{\pm \mp}}$.
Finally, the last term relates to the carrier transfer coming from the opposite valley (source term).
As Kerr rotation measures the spin/valley imbalance of the system, we model our measurements as the difference between the valley populations $\theta _K = (n_++p_+)-(n_-+p_-)$.
The initial conditions allow us to consider the two types of excitation, for instance, $n_+(0)=N/2$, $p_+(0)=N/2$, $n_-(0)=0$, and $p_-(0)=0$ for $\sigma_+$ excitation, where $N/2$ is the number of excited carriers.

The results obtained by solving our rate-equation system are in good agreement with the experimental measurements with and without magnetic fields. 
In Figure \ref{Fig4:model}a we compare our experimental and theoretical results at zero magnetic field, where valley degeneracy is still present. 
We noticed that to properly reproduce our measurements, the radiative recombination rate could not be either the shorter nor the longer decay process, and instead, good results were achieved by using decay times ($\tau$) of tens of picoseconds.
The latter agrees with previous reports of the trion lifetime for MoSe$_2$ \cite{Wang2015, Godde2016, Anghel2018, Zhang2021}.
Therefore, as our results point to a spin accumulation of holes at the long time range ($\tau_v$ = 5 $ns$), the fast decay ($\tau_c$ = 1.8 $ps$) is assigned to electron transfer at the conduction band.
This is consistent with previous reports pointing out to the fast depolarization of electrons in the conduction band when compared to the holes, and even radiative recombination of trions \cite{Hsu2015, Mai2014}.

To simulate our results at high magnetic fields (Figure \ref{Fig4:model}b) we introduce an asymmetry in the scattering times of the different considered processes (see supplementary information for details). 
Our model indicates that the main process responsible for the fast switch of spin polarization is a strong reduction of the hole inter-valley scattering through the valley with higher valence band energy.
In the right panel of Figure \ref{Fig4:model}b we show the result of the TRKR calculation just with the effect of an asymmetric scattering process at the valence band, by considering $\tau _{v_{-+}}$ = 5 ps, three orders of magnitude smaller than $\tau _{v_{+-}}$.
Meanwhile, the introduction of different scattering times for the electrons in the conduction band and different radiative recombination times at the the two valleys can help to refine the result at the short lifetimes but they have no major impact on the total decay profile. 
We point out that in both calculations in Figures \ref{Fig4:model}a and \ref{Fig4:model}b, we did not obtain a long-lived decay process, such as the ones observed in the experimental results, even considering larger scattering lifetimes for holes $\tau _v$.
Therefore, we associate this long-lived component to resident charge carriers in agreement with previous reports \cite{Schwemmer2017, Ersfeld2019, Anghel2018, Hsu2015}. 
In Figure \ref{Fig4:model}c, we illustrate the obtained spin dynamics for positive magnetic field and an optical excitation of the $K_-$ valley.

Our results show that optically-generated spins in TMDs can be very effectively controlled by magnetic fields, which is explained by a strong field-induced asymmetry on the intervalley scattering rates.
We envision that the external magnetic field used in our experiments could be replaced by magnetic proximity to van der Waals magnets.
These heterostructures have already shown to result on a significant control over valley polarization in photoluminescence measurements \cite{Seyler2018}, and our work implies that a significant polarization of long-lived resident carriers can potentially be efficiently obtained and controlled via proximity effects, with lifetimes of several ns opposed to a few ps for excitons/trions.
We believe that this will open new directions for opto-spintronic devices based on TMDs which do not depend on the stabilization of bound electron-hole pairs, but  make use of the full potential of the long spin lifetimes offered by resident carriers in these materials.

See Supplementary Information for additional details regarding the experimental setup, photoluminescence measurements with magnetic field, TRKR data at different wavelengths with an external magnetic field, TRDR, and further discussion of the rate-equation model.


We thank J. G. Holstein, H. de Vries, F. van der Velde, H. Adema, and A. Joshua for their technical support.
This work was supported by the Dutch Research Council (NWO—STU.019.014), the Zernike Institute for Advanced Materials, and the Brazilian funding agencies CNPq, FAPEMIG and the Coordenação de Aperfeiçoamento de Pessoal de Nível Superior - Brasil (CAPES) - Project code 88887.476316/2020-00.






\nocite{*}

\bibliography{main}

\begin{thebibliography}{37}%
\makeatletter
\providecommand \@ifxundefined [1]{%
 \@ifx{#1\undefined}
}%
\providecommand \@ifnum [1]{%
 \ifnum #1\expandafter \@firstoftwo
 \else \expandafter \@secondoftwo
 \fi
}%
\providecommand \@ifx [1]{%
 \ifx #1\expandafter \@firstoftwo
 \else \expandafter \@secondoftwo
 \fi
}%
\providecommand \natexlab [1]{#1}%
\providecommand \enquote  [1]{``#1''}%
\providecommand \bibnamefont  [1]{#1}%
\providecommand \bibfnamefont [1]{#1}%
\providecommand \citenamefont [1]{#1}%
\providecommand \href@noop [0]{\@secondoftwo}%
\providecommand \href [0]{\begingroup \@sanitize@url \@href}%
\providecommand \@href[1]{\@@startlink{#1}\@@href}%
\providecommand \@@href[1]{\endgroup#1\@@endlink}%
\providecommand \@sanitize@url [0]{\catcode `\\12\catcode `\$12\catcode
  `\&12\catcode `\#12\catcode `\^12\catcode `\_12\catcode `\%12\relax}%
\providecommand \@@startlink[1]{}%
\providecommand \@@endlink[0]{}%
\providecommand \url  [0]{\begingroup\@sanitize@url \@url }%
\providecommand \@url [1]{\endgroup\@href {#1}{\urlprefix }}%
\providecommand \urlprefix  [0]{URL }%
\providecommand \Eprint [0]{\href }%
\providecommand \doibase [0]{https://doi.org/}%
\providecommand \selectlanguage [0]{\@gobble}%
\providecommand \bibinfo  [0]{\@secondoftwo}%
\providecommand \bibfield  [0]{\@secondoftwo}%
\providecommand \translation [1]{[#1]}%
\providecommand \BibitemOpen [0]{}%
\providecommand \bibitemStop [0]{}%
\providecommand \bibitemNoStop [0]{.\EOS\space}%
\providecommand \EOS [0]{\spacefactor3000\relax}%
\providecommand \BibitemShut  [1]{\csname bibitem#1\endcsname}%
\let\auto@bib@innerbib\@empty
\bibitem [{\citenamefont {Mak}\ \emph {et~al.}(2012)\citenamefont {Mak},
  \citenamefont {He}, \citenamefont {Shan},\ and\ \citenamefont
  {Heinz}}]{Mak2012}%
  \BibitemOpen
  \bibfield  {author} {\bibinfo {author} {\bibfnamefont {K.~F.}\ \bibnamefont
  {Mak}}, \bibinfo {author} {\bibfnamefont {K.}~\bibnamefont {He}}, \bibinfo
  {author} {\bibfnamefont {J.}~\bibnamefont {Shan}},\ and\ \bibinfo {author}
  {\bibfnamefont {T.~F.}\ \bibnamefont {Heinz}},\ }\bibfield  {title} {\bibinfo
  {title} {Control of valley polarization in monolayer {M}o{S}$_2$ by optical
  helicity},\ }\href@noop {} {\bibfield  {journal} {\bibinfo  {journal} {Nat.
  Nanotech.}\ }\textbf {\bibinfo {volume} {7}},\ \bibinfo {pages} {494}
  (\bibinfo {year} {2012})}\BibitemShut {NoStop}%
\bibitem [{\citenamefont {Zeng}\ \emph {et~al.}(2012)\citenamefont {Zeng},
  \citenamefont {Dai}, \citenamefont {Yao}, \citenamefont {Xiao},\ and\
  \citenamefont {Cui}}]{Zeng2012}%
  \BibitemOpen
  \bibfield  {author} {\bibinfo {author} {\bibfnamefont {H.}~\bibnamefont
  {Zeng}}, \bibinfo {author} {\bibfnamefont {J.}~\bibnamefont {Dai}}, \bibinfo
  {author} {\bibfnamefont {W.}~\bibnamefont {Yao}}, \bibinfo {author}
  {\bibfnamefont {D.}~\bibnamefont {Xiao}},\ and\ \bibinfo {author}
  {\bibfnamefont {X.}~\bibnamefont {Cui}},\ }\bibfield  {title} {\bibinfo
  {title} {Valley polarization in {M}o{S}$_2$ monolayers by optical pumping},\
  }\href@noop {} {\bibfield  {journal} {\bibinfo  {journal} {Nat. Nanotech.}\
  }\textbf {\bibinfo {volume} {7}},\ \bibinfo {pages} {490} (\bibinfo {year}
  {2012})}\BibitemShut {NoStop}%
\bibitem [{\citenamefont {Xu}\ \emph {et~al.}(2014)\citenamefont {Xu},
  \citenamefont {Yao}, \citenamefont {Xiao},\ and\ \citenamefont
  {Heinz}}]{Xu2014}%
  \BibitemOpen
  \bibfield  {author} {\bibinfo {author} {\bibfnamefont {X.}~\bibnamefont
  {Xu}}, \bibinfo {author} {\bibfnamefont {W.}~\bibnamefont {Yao}}, \bibinfo
  {author} {\bibfnamefont {D.}~\bibnamefont {Xiao}},\ and\ \bibinfo {author}
  {\bibfnamefont {T.~F.}\ \bibnamefont {Heinz}},\ }\bibfield  {title} {\bibinfo
  {title} {Spin and pseudospins in layered transition metal dichalcogenides},\
  }\href@noop {} {\bibfield  {journal} {\bibinfo  {journal} {Nat. Phys.}\
  }\textbf {\bibinfo {volume} {10}},\ \bibinfo {pages} {343} (\bibinfo {year}
  {2014})}\BibitemShut {NoStop}%
\bibitem [{\citenamefont {Mak}\ and\ \citenamefont {Shan}(2016)}]{Mak2016}%
  \BibitemOpen
  \bibfield  {author} {\bibinfo {author} {\bibfnamefont {K.~F.}\ \bibnamefont
  {Mak}}\ and\ \bibinfo {author} {\bibfnamefont {J.}~\bibnamefont {Shan}},\
  }\bibfield  {title} {\bibinfo {title} {Photonics and optoelectronics of 2{D}
  semiconductor transition metal dichalcogenides},\ }\href@noop {} {\bibfield
  {journal} {\bibinfo  {journal} {Nat. Photon.}\ }\textbf {\bibinfo {volume}
  {10}},\ \bibinfo {pages} {216} (\bibinfo {year} {2016})}\BibitemShut
  {NoStop}%
\bibitem [{\citenamefont {Liu}\ \emph {et~al.}(2016)\citenamefont {Liu},
  \citenamefont {Weiss}, \citenamefont {Duan}, \citenamefont {Cheng},
  \citenamefont {Huang},\ and\ \citenamefont {Duan}}]{Liu2016}%
  \BibitemOpen
  \bibfield  {author} {\bibinfo {author} {\bibfnamefont {Y.}~\bibnamefont
  {Liu}}, \bibinfo {author} {\bibfnamefont {N.~O.}\ \bibnamefont {Weiss}},
  \bibinfo {author} {\bibfnamefont {X.}~\bibnamefont {Duan}}, \bibinfo {author}
  {\bibfnamefont {H.~C.}\ \bibnamefont {Cheng}}, \bibinfo {author}
  {\bibfnamefont {Y.}~\bibnamefont {Huang}},\ and\ \bibinfo {author}
  {\bibfnamefont {X.}~\bibnamefont {Duan}},\ }\bibfield  {title} {\bibinfo
  {title} {Van der {W}aals heterostructures and devices},\ }\href@noop {}
  {\bibfield  {journal} {\bibinfo  {journal} {Nat. Rev. Mater.}\ }\textbf
  {\bibinfo {volume} {1}},\ \bibinfo {pages} {16042} (\bibinfo {year}
  {2016})}\BibitemShut {NoStop}%
\bibitem [{\citenamefont {Schaibley}\ \emph {et~al.}(2016)\citenamefont
  {Schaibley}, \citenamefont {Yu}, \citenamefont {Clark}, \citenamefont
  {Rivera}, \citenamefont {Ross}, \citenamefont {Seyler}, \citenamefont {Yao},\
  and\ \citenamefont {Xu}}]{Schaibley2016}%
  \BibitemOpen
  \bibfield  {author} {\bibinfo {author} {\bibfnamefont {J.~R.}\ \bibnamefont
  {Schaibley}}, \bibinfo {author} {\bibfnamefont {H.}~\bibnamefont {Yu}},
  \bibinfo {author} {\bibfnamefont {G.}~\bibnamefont {Clark}}, \bibinfo
  {author} {\bibfnamefont {P.}~\bibnamefont {Rivera}}, \bibinfo {author}
  {\bibfnamefont {J.~S.}\ \bibnamefont {Ross}}, \bibinfo {author}
  {\bibfnamefont {K.~L.}\ \bibnamefont {Seyler}}, \bibinfo {author}
  {\bibfnamefont {W.}~\bibnamefont {Yao}},\ and\ \bibinfo {author}
  {\bibfnamefont {X.}~\bibnamefont {Xu}},\ }\bibfield  {title} {\bibinfo
  {title} {Valleytronics in 2{D} materials},\ }\href@noop {} {\bibfield
  {journal} {\bibinfo  {journal} {Nat. Rev. Mater.}\ }\textbf {\bibinfo
  {volume} {1}},\ \bibinfo {pages} {16055} (\bibinfo {year}
  {2016})}\BibitemShut {NoStop}%
\bibitem [{\citenamefont {Zhong}\ \emph {et~al.}(2017)\citenamefont {Zhong},
  \citenamefont {Seyler}, \citenamefont {Linpeng}, \citenamefont {Cheng},
  \citenamefont {Sivadas}, \citenamefont {Huang}, \citenamefont {Schmidgall},
  \citenamefont {Taniguchi}, \citenamefont {Watanabe}, \citenamefont {McGuire},
  \citenamefont {Yao}, \citenamefont {Xiao}, \citenamefont {Fu},\ and\
  \citenamefont {Xu}}]{Zhong2017}%
  \BibitemOpen
  \bibfield  {author} {\bibinfo {author} {\bibfnamefont {D.}~\bibnamefont
  {Zhong}}, \bibinfo {author} {\bibfnamefont {K.~L.}\ \bibnamefont {Seyler}},
  \bibinfo {author} {\bibfnamefont {X.}~\bibnamefont {Linpeng}}, \bibinfo
  {author} {\bibfnamefont {R.}~\bibnamefont {Cheng}}, \bibinfo {author}
  {\bibfnamefont {N.}~\bibnamefont {Sivadas}}, \bibinfo {author} {\bibfnamefont
  {B.}~\bibnamefont {Huang}}, \bibinfo {author} {\bibfnamefont
  {E.}~\bibnamefont {Schmidgall}}, \bibinfo {author} {\bibfnamefont
  {T.}~\bibnamefont {Taniguchi}}, \bibinfo {author} {\bibfnamefont
  {K.}~\bibnamefont {Watanabe}}, \bibinfo {author} {\bibfnamefont {M.~A.}\
  \bibnamefont {McGuire}}, \bibinfo {author} {\bibfnamefont {W.}~\bibnamefont
  {Yao}}, \bibinfo {author} {\bibfnamefont {D.}~\bibnamefont {Xiao}}, \bibinfo
  {author} {\bibfnamefont {K.-M.~C.}\ \bibnamefont {Fu}},\ and\ \bibinfo
  {author} {\bibfnamefont {X.}~\bibnamefont {Xu}},\ }\bibfield  {title}
  {\bibinfo {title} {Van der {W}aals engineering of ferromagnetic semiconductor
  heterostructures for spin and valleytronics},\ }\href@noop {} {\bibfield
  {journal} {\bibinfo  {journal} {Sci. Adv.}\ }\textbf {\bibinfo {volume}
  {3}},\ \bibinfo {pages} {e1603113} (\bibinfo {year} {2017})}\BibitemShut
  {NoStop}%
\bibitem [{\citenamefont {Luo}\ \emph {et~al.}(2017)\citenamefont {Luo},
  \citenamefont {Xu}, \citenamefont {Zhu}, \citenamefont {Wu}, \citenamefont
  {McCormick}, \citenamefont {Zhan}, \citenamefont {Neupane},\ and\
  \citenamefont {Kawakami}}]{Luo2017}%
  \BibitemOpen
  \bibfield  {author} {\bibinfo {author} {\bibfnamefont {Y.~K.}\ \bibnamefont
  {Luo}}, \bibinfo {author} {\bibfnamefont {J.}~\bibnamefont {Xu}}, \bibinfo
  {author} {\bibfnamefont {T.}~\bibnamefont {Zhu}}, \bibinfo {author}
  {\bibfnamefont {G.}~\bibnamefont {Wu}}, \bibinfo {author} {\bibfnamefont
  {E.~J.}\ \bibnamefont {McCormick}}, \bibinfo {author} {\bibfnamefont
  {W.}~\bibnamefont {Zhan}}, \bibinfo {author} {\bibfnamefont {M.~R.}\
  \bibnamefont {Neupane}},\ and\ \bibinfo {author} {\bibfnamefont {R.~K.}\
  \bibnamefont {Kawakami}},\ }\bibfield  {title} {\bibinfo {title}
  {Opto-valleytronic spin injection in monolayer {M}o{S}$_2$/few-layer graphene
  hybrid spin valves},\ }\href@noop {} {\bibfield  {journal} {\bibinfo
  {journal} {Nano Lett.}\ }\textbf {\bibinfo {volume} {17}},\ \bibinfo {pages}
  {3877} (\bibinfo {year} {2017})}\BibitemShut {NoStop}%
\bibitem [{\citenamefont {MacNeill}\ \emph {et~al.}(2015)\citenamefont
  {MacNeill}, \citenamefont {Heikes}, \citenamefont {Mak}, \citenamefont
  {Anderson}, \citenamefont {Korm\'anyos}, \citenamefont {Z\'olyomi},
  \citenamefont {Park},\ and\ \citenamefont {Ralph}}]{Macneill2015}%
  \BibitemOpen
  \bibfield  {author} {\bibinfo {author} {\bibfnamefont {D.}~\bibnamefont
  {MacNeill}}, \bibinfo {author} {\bibfnamefont {C.}~\bibnamefont {Heikes}},
  \bibinfo {author} {\bibfnamefont {K.~F.}\ \bibnamefont {Mak}}, \bibinfo
  {author} {\bibfnamefont {Z.}~\bibnamefont {Anderson}}, \bibinfo {author}
  {\bibfnamefont {A.}~\bibnamefont {Korm\'anyos}}, \bibinfo {author}
  {\bibfnamefont {V.}~\bibnamefont {Z\'olyomi}}, \bibinfo {author}
  {\bibfnamefont {J.}~\bibnamefont {Park}},\ and\ \bibinfo {author}
  {\bibfnamefont {D.~C.}\ \bibnamefont {Ralph}},\ }\bibfield  {title} {\bibinfo
  {title} {Breaking of valley degeneracy by magnetic field in monolayer
  {M}o{S}e$_2$},\ }\href@noop {} {\bibfield  {journal} {\bibinfo  {journal}
  {Phys. Rev. Lett.}\ }\textbf {\bibinfo {volume} {114}},\ \bibinfo {pages}
  {037401} (\bibinfo {year} {2015})}\BibitemShut {NoStop}%
\bibitem [{\citenamefont {Li}\ \emph {et~al.}(2014)\citenamefont {Li},
  \citenamefont {Ludwig}, \citenamefont {Low}, \citenamefont {Chernikov},
  \citenamefont {Cui}, \citenamefont {Arefe}, \citenamefont {Kim},
  \citenamefont {van~der Zande}, \citenamefont {Rigosi}, \citenamefont {Hill},
  \citenamefont {Kim}, \citenamefont {Hone}, \citenamefont {Li}, \citenamefont
  {Smirnov},\ and\ \citenamefont {Heinz}}]{Li2014}%
  \BibitemOpen
  \bibfield  {author} {\bibinfo {author} {\bibfnamefont {Y.}~\bibnamefont
  {Li}}, \bibinfo {author} {\bibfnamefont {J.}~\bibnamefont {Ludwig}}, \bibinfo
  {author} {\bibfnamefont {T.}~\bibnamefont {Low}}, \bibinfo {author}
  {\bibfnamefont {A.}~\bibnamefont {Chernikov}}, \bibinfo {author}
  {\bibfnamefont {X.}~\bibnamefont {Cui}}, \bibinfo {author} {\bibfnamefont
  {G.}~\bibnamefont {Arefe}}, \bibinfo {author} {\bibfnamefont {Y.~D.}\
  \bibnamefont {Kim}}, \bibinfo {author} {\bibfnamefont {A.~M.}\ \bibnamefont
  {van~der Zande}}, \bibinfo {author} {\bibfnamefont {A.}~\bibnamefont
  {Rigosi}}, \bibinfo {author} {\bibfnamefont {H.~M.}\ \bibnamefont {Hill}},
  \bibinfo {author} {\bibfnamefont {S.~H.}\ \bibnamefont {Kim}}, \bibinfo
  {author} {\bibfnamefont {J.}~\bibnamefont {Hone}}, \bibinfo {author}
  {\bibfnamefont {Z.}~\bibnamefont {Li}}, \bibinfo {author} {\bibfnamefont
  {D.}~\bibnamefont {Smirnov}},\ and\ \bibinfo {author} {\bibfnamefont {T.~F.}\
  \bibnamefont {Heinz}},\ }\bibfield  {title} {\bibinfo {title} {Valley
  splitting and polarization by the {Z}eeman effect in monolayer
  {M}o{S}e$_2$},\ }\href@noop {} {\bibfield  {journal} {\bibinfo  {journal}
  {Phys. Rev. Lett.}\ }\textbf {\bibinfo {volume} {113}},\ \bibinfo {pages}
  {266804} (\bibinfo {year} {2014})}\BibitemShut {NoStop}%
\bibitem [{\citenamefont {Srivastava}\ \emph {et~al.}(2015)\citenamefont
  {Srivastava}, \citenamefont {Sidler}, \citenamefont {Allain}, \citenamefont
  {Lembke}, \citenamefont {Kis},\ and\ \citenamefont
  {Imamoğlu}}]{Srivastava2015}%
  \BibitemOpen
  \bibfield  {author} {\bibinfo {author} {\bibfnamefont {A.}~\bibnamefont
  {Srivastava}}, \bibinfo {author} {\bibfnamefont {M.}~\bibnamefont {Sidler}},
  \bibinfo {author} {\bibfnamefont {A.~V.}\ \bibnamefont {Allain}}, \bibinfo
  {author} {\bibfnamefont {D.~S.}\ \bibnamefont {Lembke}}, \bibinfo {author}
  {\bibfnamefont {A.}~\bibnamefont {Kis}},\ and\ \bibinfo {author}
  {\bibfnamefont {A.}~\bibnamefont {Imamoğlu}},\ }\bibfield  {title} {\bibinfo
  {title} {Valley {Z}eeman effect in elementary optical excitations of
  monolayer {WS}e$_2$},\ }\href@noop {} {\bibfield  {journal} {\bibinfo
  {journal} {Nat. Phys.}\ }\textbf {\bibinfo {volume} {11}},\ \bibinfo {pages}
  {141} (\bibinfo {year} {2015})}\BibitemShut {NoStop}%
\bibitem [{\citenamefont {Aivazian}\ \emph {et~al.}(2015)\citenamefont
  {Aivazian}, \citenamefont {Gong}, \citenamefont {Jones}, \citenamefont {Chu},
  \citenamefont {Yan}, \citenamefont {Mandrus}, \citenamefont {Zhang},
  \citenamefont {Cobden}, \citenamefont {Yao},\ and\ \citenamefont
  {Xu}}]{Aivazian2015}%
  \BibitemOpen
  \bibfield  {author} {\bibinfo {author} {\bibfnamefont {G.}~\bibnamefont
  {Aivazian}}, \bibinfo {author} {\bibfnamefont {Z.}~\bibnamefont {Gong}},
  \bibinfo {author} {\bibfnamefont {A.~M.}\ \bibnamefont {Jones}}, \bibinfo
  {author} {\bibfnamefont {R.~L.}\ \bibnamefont {Chu}}, \bibinfo {author}
  {\bibfnamefont {J.}~\bibnamefont {Yan}}, \bibinfo {author} {\bibfnamefont
  {D.~G.}\ \bibnamefont {Mandrus}}, \bibinfo {author} {\bibfnamefont
  {C.}~\bibnamefont {Zhang}}, \bibinfo {author} {\bibfnamefont
  {D.}~\bibnamefont {Cobden}}, \bibinfo {author} {\bibfnamefont
  {W.}~\bibnamefont {Yao}},\ and\ \bibinfo {author} {\bibfnamefont
  {X.}~\bibnamefont {Xu}},\ }\bibfield  {title} {\bibinfo {title} {Magnetic
  control of valley pseudospin in monolayer {WS}e$_2$},\ }\href@noop {}
  {\bibfield  {journal} {\bibinfo  {journal} {Nat. Phys.}\ }\textbf {\bibinfo
  {volume} {11}},\ \bibinfo {pages} {148} (\bibinfo {year} {2015})}\BibitemShut
  {NoStop}%
\bibitem [{\citenamefont {Koperski}\ \emph {et~al.}(2019)\citenamefont
  {Koperski}, \citenamefont {Molas}, \citenamefont {Arora}, \citenamefont
  {Nogajewski}, \citenamefont {Bartos}, \citenamefont {Wyzula}, \citenamefont
  {Vaclavkova}, \citenamefont {Kossacki},\ and\ \citenamefont
  {Potemski}}]{Koperski2019}%
  \BibitemOpen
  \bibfield  {author} {\bibinfo {author} {\bibfnamefont {M.}~\bibnamefont
  {Koperski}}, \bibinfo {author} {\bibfnamefont {M.~R.}\ \bibnamefont {Molas}},
  \bibinfo {author} {\bibfnamefont {A.}~\bibnamefont {Arora}}, \bibinfo
  {author} {\bibfnamefont {K.}~\bibnamefont {Nogajewski}}, \bibinfo {author}
  {\bibfnamefont {M.}~\bibnamefont {Bartos}}, \bibinfo {author} {\bibfnamefont
  {J.}~\bibnamefont {Wyzula}}, \bibinfo {author} {\bibfnamefont
  {D.}~\bibnamefont {Vaclavkova}}, \bibinfo {author} {\bibfnamefont
  {P.}~\bibnamefont {Kossacki}},\ and\ \bibinfo {author} {\bibfnamefont
  {M.}~\bibnamefont {Potemski}},\ }\bibfield  {title} {\bibinfo {title}
  {Orbital, spin and valley contributions to {Z}eeman splitting of excitonic
  resonances in {M}o{S}e$_2$, {WS}e$_2$ and {WS}$_2$ monolayers},\ }\href@noop
  {} {\bibfield  {journal} {\bibinfo  {journal} {2D Mater.}\ }\textbf {\bibinfo
  {volume} {6}},\ \bibinfo {pages} {015001} (\bibinfo {year}
  {2019})}\BibitemShut {NoStop}%
\bibitem [{\citenamefont {Wang}\ \emph
  {et~al.}(2015{\natexlab{a}})\citenamefont {Wang}, \citenamefont {Bouet},
  \citenamefont {Glazov}, \citenamefont {Amand}, \citenamefont {Ivchenko},
  \citenamefont {Palleau}, \citenamefont {Marie},\ and\ \citenamefont
  {Urbaszek}}]{Wang20152D}%
  \BibitemOpen
  \bibfield  {author} {\bibinfo {author} {\bibfnamefont {G.}~\bibnamefont
  {Wang}}, \bibinfo {author} {\bibfnamefont {L.}~\bibnamefont {Bouet}},
  \bibinfo {author} {\bibfnamefont {M.~M.}\ \bibnamefont {Glazov}}, \bibinfo
  {author} {\bibfnamefont {T.}~\bibnamefont {Amand}}, \bibinfo {author}
  {\bibfnamefont {E.~L.}\ \bibnamefont {Ivchenko}}, \bibinfo {author}
  {\bibfnamefont {E.}~\bibnamefont {Palleau}}, \bibinfo {author} {\bibfnamefont
  {X.}~\bibnamefont {Marie}},\ and\ \bibinfo {author} {\bibfnamefont
  {B.}~\bibnamefont {Urbaszek}},\ }\bibfield  {title} {\bibinfo {title}
  {Magneto-optics in transition metal diselenide monolayers},\ }\href@noop {}
  {\bibfield  {journal} {\bibinfo  {journal} {2D Mater.}\ }\textbf {\bibinfo
  {volume} {2}},\ \bibinfo {pages} {034002} (\bibinfo {year}
  {2015}{\natexlab{a}})}\BibitemShut {NoStop}%
\bibitem [{\citenamefont {Rybkovskiy}\ \emph {et~al.}(2017)\citenamefont
  {Rybkovskiy}, \citenamefont {Gerber},\ and\ \citenamefont
  {Durnev}}]{Rybkovskiy2017}%
  \BibitemOpen
  \bibfield  {author} {\bibinfo {author} {\bibfnamefont {D.~V.}\ \bibnamefont
  {Rybkovskiy}}, \bibinfo {author} {\bibfnamefont {I.~C.}\ \bibnamefont
  {Gerber}},\ and\ \bibinfo {author} {\bibfnamefont {M.~V.}\ \bibnamefont
  {Durnev}},\ }\bibfield  {title} {\bibinfo {title} {Atomically inspired
  $k\ifmmode\cdot\else\textperiodcentered\fi{}p$ approach and valley {Z}eeman
  effect in transition metal dichalcogenide monolayers},\ }\href@noop {}
  {\bibfield  {journal} {\bibinfo  {journal} {Phys. Rev. B}\ }\textbf {\bibinfo
  {volume} {95}},\ \bibinfo {pages} {155406} (\bibinfo {year}
  {2017})}\BibitemShut {NoStop}%
\bibitem [{\citenamefont {Wo\ifmmode~\acute{z}\else \'{z}\fi{}niak}\ \emph
  {et~al.}(2020)\citenamefont {Wo\ifmmode~\acute{z}\else \'{z}\fi{}niak},
  \citenamefont {Faria~Junior}, \citenamefont {Seifert}, \citenamefont
  {Chaves},\ and\ \citenamefont {Kunstmann}}]{Wozniak2020}%
  \BibitemOpen
  \bibfield  {author} {\bibinfo {author} {\bibfnamefont {T.}~\bibnamefont
  {Wo\ifmmode~\acute{z}\else \'{z}\fi{}niak}}, \bibinfo {author} {\bibfnamefont
  {P.~E.}\ \bibnamefont {Faria~Junior}}, \bibinfo {author} {\bibfnamefont
  {G.}~\bibnamefont {Seifert}}, \bibinfo {author} {\bibfnamefont
  {A.}~\bibnamefont {Chaves}},\ and\ \bibinfo {author} {\bibfnamefont
  {J.}~\bibnamefont {Kunstmann}},\ }\bibfield  {title} {\bibinfo {title}
  {Exciton $g$ factors of van der {W}aals heterostructures from
  first-principles calculations},\ }\href@noop {} {\bibfield  {journal}
  {\bibinfo  {journal} {Phys. Rev. B}\ }\textbf {\bibinfo {volume} {101}},\
  \bibinfo {pages} {235408} (\bibinfo {year} {2020})}\BibitemShut {NoStop}%
\bibitem [{\citenamefont {Wang}\ \emph
  {et~al.}(2015{\natexlab{b}})\citenamefont {Wang}, \citenamefont {Palleau},
  \citenamefont {Amand}, \citenamefont {Tongay}, \citenamefont {Marie},\ and\
  \citenamefont {Urbaszek}}]{Wang2015}%
  \BibitemOpen
  \bibfield  {author} {\bibinfo {author} {\bibfnamefont {G.}~\bibnamefont
  {Wang}}, \bibinfo {author} {\bibfnamefont {E.}~\bibnamefont {Palleau}},
  \bibinfo {author} {\bibfnamefont {T.}~\bibnamefont {Amand}}, \bibinfo
  {author} {\bibfnamefont {S.}~\bibnamefont {Tongay}}, \bibinfo {author}
  {\bibfnamefont {X.}~\bibnamefont {Marie}},\ and\ \bibinfo {author}
  {\bibfnamefont {B.}~\bibnamefont {Urbaszek}},\ }\bibfield  {title} {\bibinfo
  {title} {Polarization and time-resolved photoluminescence spectroscopy of
  excitons in {M}o{S}e$_2$ monolayers},\ }\href@noop {} {\bibfield  {journal}
  {\bibinfo  {journal} {Appl. Phys. Lett.}\ }\textbf {\bibinfo {volume}
  {106}},\ \bibinfo {pages} {112101} (\bibinfo {year}
  {2015}{\natexlab{b}})}\BibitemShut {NoStop}%
\bibitem [{\citenamefont {Godde}\ \emph {et~al.}(2016)\citenamefont {Godde},
  \citenamefont {Schmidt}, \citenamefont {Schmutzler}, \citenamefont
  {A\ss{}mann}, \citenamefont {Debus}, \citenamefont {Withers}, \citenamefont
  {Alexeev}, \citenamefont {Del Pozo-Zamudio}, \citenamefont {Skrypka},
  \citenamefont {Novoselov}, \citenamefont {Bayer},\ and\ \citenamefont
  {Tartakovskii}}]{Godde2016}%
  \BibitemOpen
  \bibfield  {author} {\bibinfo {author} {\bibfnamefont {T.}~\bibnamefont
  {Godde}}, \bibinfo {author} {\bibfnamefont {D.}~\bibnamefont {Schmidt}},
  \bibinfo {author} {\bibfnamefont {J.}~\bibnamefont {Schmutzler}}, \bibinfo
  {author} {\bibfnamefont {M.}~\bibnamefont {A\ss{}mann}}, \bibinfo {author}
  {\bibfnamefont {J.}~\bibnamefont {Debus}}, \bibinfo {author} {\bibfnamefont
  {F.}~\bibnamefont {Withers}}, \bibinfo {author} {\bibfnamefont {E.~M.}\
  \bibnamefont {Alexeev}}, \bibinfo {author} {\bibfnamefont {O.}~\bibnamefont
  {Del Pozo-Zamudio}}, \bibinfo {author} {\bibfnamefont {O.~V.}\ \bibnamefont
  {Skrypka}}, \bibinfo {author} {\bibfnamefont {K.~S.}\ \bibnamefont
  {Novoselov}}, \bibinfo {author} {\bibfnamefont {M.}~\bibnamefont {Bayer}},\
  and\ \bibinfo {author} {\bibfnamefont {A.~I.}\ \bibnamefont {Tartakovskii}},\
  }\bibfield  {title} {\bibinfo {title} {Exciton and trion dynamics in
  atomically thin {M}o{S}e$_2$ and {WS}e$_2$: Effect of localization},\
  }\href@noop {} {\bibfield  {journal} {\bibinfo  {journal} {Phys. Rev. B}\
  }\textbf {\bibinfo {volume} {94}},\ \bibinfo {pages} {165301} (\bibinfo
  {year} {2016})}\BibitemShut {NoStop}%
\bibitem [{\citenamefont {Wang}\ and\ \citenamefont {Ma}(2020)}]{Wang2020}%
  \BibitemOpen
  \bibfield  {author} {\bibinfo {author} {\bibfnamefont {W.}~\bibnamefont
  {Wang}}\ and\ \bibinfo {author} {\bibfnamefont {X.}~\bibnamefont {Ma}},\
  }\bibfield  {title} {\bibinfo {title} {Strain-induced trapping of indirect
  excitons in {M}o{S}e$_2$/{WS}e$_2$ heterostructures},\ }\href@noop {}
  {\bibfield  {journal} {\bibinfo  {journal} {ACS Photonics}\ }\textbf
  {\bibinfo {volume} {7}},\ \bibinfo {pages} {2460} (\bibinfo {year}
  {2020})}\BibitemShut {NoStop}%
\bibitem [{\citenamefont {Kumar}\ \emph
  {et~al.}(2014{\natexlab{a}})\citenamefont {Kumar}, \citenamefont {He},
  \citenamefont {He}, \citenamefont {Wang},\ and\ \citenamefont
  {Zhao}}]{Kumar2014}%
  \BibitemOpen
  \bibfield  {author} {\bibinfo {author} {\bibfnamefont {N.}~\bibnamefont
  {Kumar}}, \bibinfo {author} {\bibfnamefont {J.}~\bibnamefont {He}}, \bibinfo
  {author} {\bibfnamefont {D.}~\bibnamefont {He}}, \bibinfo {author}
  {\bibfnamefont {Y.}~\bibnamefont {Wang}},\ and\ \bibinfo {author}
  {\bibfnamefont {H.}~\bibnamefont {Zhao}},\ }\bibfield  {title} {\bibinfo
  {title} {Valley and spin dynamics in {M}o{S}e$_2$ two-dimensional crystals},\
  }\href@noop {} {\bibfield  {journal} {\bibinfo  {journal} {Nanoscale}\
  }\textbf {\bibinfo {volume} {6}},\ \bibinfo {pages} {12690} (\bibinfo {year}
  {2014}{\natexlab{a}})}\BibitemShut {NoStop}%
\bibitem [{\citenamefont {Kumar}\ \emph
  {et~al.}(2014{\natexlab{b}})\citenamefont {Kumar}, \citenamefont {Cui},
  \citenamefont {Ceballos}, \citenamefont {He}, \citenamefont {Wang},\ and\
  \citenamefont {Zhao}}]{Kumar2014prb}%
  \BibitemOpen
  \bibfield  {author} {\bibinfo {author} {\bibfnamefont {N.}~\bibnamefont
  {Kumar}}, \bibinfo {author} {\bibfnamefont {Q.}~\bibnamefont {Cui}}, \bibinfo
  {author} {\bibfnamefont {F.}~\bibnamefont {Ceballos}}, \bibinfo {author}
  {\bibfnamefont {D.}~\bibnamefont {He}}, \bibinfo {author} {\bibfnamefont
  {Y.}~\bibnamefont {Wang}},\ and\ \bibinfo {author} {\bibfnamefont
  {H.}~\bibnamefont {Zhao}},\ }\bibfield  {title} {\bibinfo {title}
  {Exciton-exciton annihilation in {M}o{S}e$_2$ monolayers},\ }\href@noop {}
  {\bibfield  {journal} {\bibinfo  {journal} {Phys. Rev. B}\ }\textbf {\bibinfo
  {volume} {89}},\ \bibinfo {pages} {125427} (\bibinfo {year}
  {2014}{\natexlab{b}})}\BibitemShut {NoStop}%
\bibitem [{\citenamefont {Ye}\ \emph {et~al.}(2018)\citenamefont {Ye},
  \citenamefont {Yan}, \citenamefont {Niu}, \citenamefont {Li},\ and\
  \citenamefont {Zhang}}]{Ye2018}%
  \BibitemOpen
  \bibfield  {author} {\bibinfo {author} {\bibfnamefont {J.}~\bibnamefont
  {Ye}}, \bibinfo {author} {\bibfnamefont {T.}~\bibnamefont {Yan}}, \bibinfo
  {author} {\bibfnamefont {B.}~\bibnamefont {Niu}}, \bibinfo {author}
  {\bibfnamefont {Y.}~\bibnamefont {Li}},\ and\ \bibinfo {author}
  {\bibfnamefont {X.}~\bibnamefont {Zhang}},\ }\bibfield  {title} {\bibinfo
  {title} {Nonlinear dynamics of trions under strong optical excitation in
  monolayer {M}o{S}e$_2$},\ }\href@noop {} {\bibfield  {journal} {\bibinfo
  {journal} {Sci. Rep.}\ }\textbf {\bibinfo {volume} {8}},\ \bibinfo {pages}
  {2389} (\bibinfo {year} {2018})}\BibitemShut {NoStop}%
\bibitem [{\citenamefont {Zhu}\ \emph {et~al.}(2014)\citenamefont {Zhu},
  \citenamefont {Zhang}, \citenamefont {Glazov}, \citenamefont {Urbaszek},
  \citenamefont {Amand}, \citenamefont {Ji}, \citenamefont {Liu},\ and\
  \citenamefont {Marie}}]{Zhu2014}%
  \BibitemOpen
  \bibfield  {author} {\bibinfo {author} {\bibfnamefont {C.~R.}\ \bibnamefont
  {Zhu}}, \bibinfo {author} {\bibfnamefont {K.}~\bibnamefont {Zhang}}, \bibinfo
  {author} {\bibfnamefont {M.}~\bibnamefont {Glazov}}, \bibinfo {author}
  {\bibfnamefont {B.}~\bibnamefont {Urbaszek}}, \bibinfo {author}
  {\bibfnamefont {T.}~\bibnamefont {Amand}}, \bibinfo {author} {\bibfnamefont
  {Z.~W.}\ \bibnamefont {Ji}}, \bibinfo {author} {\bibfnamefont {B.~L.}\
  \bibnamefont {Liu}},\ and\ \bibinfo {author} {\bibfnamefont {X.}~\bibnamefont
  {Marie}},\ }\bibfield  {title} {\bibinfo {title} {Exciton valley dynamics
  probed by {K}err rotation in {WS}e$_2$ monolayers},\ }\href@noop {}
  {\bibfield  {journal} {\bibinfo  {journal} {Phys. Rev. B}\ }\textbf {\bibinfo
  {volume} {90}},\ \bibinfo {pages} {161302} (\bibinfo {year}
  {2014})}\BibitemShut {NoStop}%
\bibitem [{\citenamefont {Hsu}\ \emph {et~al.}(2015)\citenamefont {Hsu},
  \citenamefont {Chen}, \citenamefont {Chen}, \citenamefont {Liu},
  \citenamefont {Hou}, \citenamefont {Li},\ and\ \citenamefont
  {Chang}}]{Hsu2015}%
  \BibitemOpen
  \bibfield  {author} {\bibinfo {author} {\bibfnamefont {W.~T.}\ \bibnamefont
  {Hsu}}, \bibinfo {author} {\bibfnamefont {Y.~L.}\ \bibnamefont {Chen}},
  \bibinfo {author} {\bibfnamefont {C.~H.}\ \bibnamefont {Chen}}, \bibinfo
  {author} {\bibfnamefont {P.~S.}\ \bibnamefont {Liu}}, \bibinfo {author}
  {\bibfnamefont {T.~H.}\ \bibnamefont {Hou}}, \bibinfo {author} {\bibfnamefont
  {L.~J.}\ \bibnamefont {Li}},\ and\ \bibinfo {author} {\bibfnamefont {W.~H.}\
  \bibnamefont {Chang}},\ }\bibfield  {title} {\bibinfo {title} {Optically
  initialized robust valley-polarized holes in monolayer {WS}e$_2$},\
  }\href@noop {} {\bibfield  {journal} {\bibinfo  {journal} {Nat. Commun.}\
  }\textbf {\bibinfo {volume} {6}},\ \bibinfo {pages} {8963} (\bibinfo {year}
  {2015})}\BibitemShut {NoStop}%
\bibitem [{\citenamefont {Guimar\~aes}\ and\ \citenamefont
  {Koopmans}(2018)}]{Guimaraes2018}%
  \BibitemOpen
  \bibfield  {author} {\bibinfo {author} {\bibfnamefont {M.~H.~D.}\
  \bibnamefont {Guimar\~aes}}\ and\ \bibinfo {author} {\bibfnamefont
  {B.}~\bibnamefont {Koopmans}},\ }\bibfield  {title} {\bibinfo {title} {Spin
  accumulation and dynamics in inversion-symmetric van der {W}aals crystals},\
  }\href@noop {} {\bibfield  {journal} {\bibinfo  {journal} {Phys. Rev. Lett.}\
  }\textbf {\bibinfo {volume} {120}},\ \bibinfo {pages} {266801} (\bibinfo
  {year} {2018})}\BibitemShut {NoStop}%
\bibitem [{\citenamefont {Anghel}\ \emph {et~al.}(2018)\citenamefont {Anghel},
  \citenamefont {Passmann}, \citenamefont {Ruppert}, \citenamefont {Bristow},\
  and\ \citenamefont {Betz}}]{Anghel2018}%
  \BibitemOpen
  \bibfield  {author} {\bibinfo {author} {\bibfnamefont {S.}~\bibnamefont
  {Anghel}}, \bibinfo {author} {\bibfnamefont {F.}~\bibnamefont {Passmann}},
  \bibinfo {author} {\bibfnamefont {C.}~\bibnamefont {Ruppert}}, \bibinfo
  {author} {\bibfnamefont {A.~D.}\ \bibnamefont {Bristow}},\ and\ \bibinfo
  {author} {\bibfnamefont {M.}~\bibnamefont {Betz}},\ }\bibfield  {title}
  {\bibinfo {title} {Coupled exciton-trion spin dynamics in a {M}o{S}e$_2$
  monolayer},\ }\href@noop {} {\bibfield  {journal} {\bibinfo  {journal} {2D
  Mater.}\ }\textbf {\bibinfo {volume} {5}},\ \bibinfo {pages} {045024}
  (\bibinfo {year} {2018})}\BibitemShut {NoStop}%
\bibitem [{\citenamefont {Ersfeld}\ \emph {et~al.}(2019)\citenamefont
  {Ersfeld}, \citenamefont {Volmer}, \citenamefont {Melo}, \citenamefont
  {Winter}, \citenamefont {Heithoff}, \citenamefont {Zanolli}, \citenamefont
  {Stampfer}, \citenamefont {Verstraete},\ and\ \citenamefont
  {Beschoten}}]{Ersfeld2019}%
  \BibitemOpen
  \bibfield  {author} {\bibinfo {author} {\bibfnamefont {M.}~\bibnamefont
  {Ersfeld}}, \bibinfo {author} {\bibfnamefont {F.}~\bibnamefont {Volmer}},
  \bibinfo {author} {\bibfnamefont {P.~M. M.~D.}\ \bibnamefont {Melo}},
  \bibinfo {author} {\bibfnamefont {R.~D.}\ \bibnamefont {Winter}}, \bibinfo
  {author} {\bibfnamefont {M.}~\bibnamefont {Heithoff}}, \bibinfo {author}
  {\bibfnamefont {Z.}~\bibnamefont {Zanolli}}, \bibinfo {author} {\bibfnamefont
  {C.}~\bibnamefont {Stampfer}}, \bibinfo {author} {\bibfnamefont {M.~J.}\
  \bibnamefont {Verstraete}},\ and\ \bibinfo {author} {\bibfnamefont
  {B.}~\bibnamefont {Beschoten}},\ }\bibfield  {title} {\bibinfo {title} {Spin
  states protected from intrinsic electron-phonon coupling reaching 100 ns
  lifetime at room temperature in {M}o{S}e$_2$},\ }\href@noop {} {\bibfield
  {journal} {\bibinfo  {journal} {Nano Lett.}\ }\textbf {\bibinfo {volume}
  {19}},\ \bibinfo {pages} {4083} (\bibinfo {year} {2019})}\BibitemShut
  {NoStop}%
\bibitem [{\citenamefont {Li}\ \emph {et~al.}(2020)\citenamefont {Li},
  \citenamefont {Wei}, \citenamefont {Ye}, \citenamefont {Zhai}, \citenamefont
  {Wang},\ and\ \citenamefont {Zhang}}]{Li2020}%
  \BibitemOpen
  \bibfield  {author} {\bibinfo {author} {\bibfnamefont {Y.}~\bibnamefont
  {Li}}, \bibinfo {author} {\bibfnamefont {X.}~\bibnamefont {Wei}}, \bibinfo
  {author} {\bibfnamefont {J.}~\bibnamefont {Ye}}, \bibinfo {author}
  {\bibfnamefont {G.}~\bibnamefont {Zhai}}, \bibinfo {author} {\bibfnamefont
  {K.}~\bibnamefont {Wang}},\ and\ \bibinfo {author} {\bibfnamefont
  {X.}~\bibnamefont {Zhang}},\ }\bibfield  {title} {\bibinfo {title}
  {Gate-controlled spin relaxation in bulk {WS}e$_2$ flakes},\ }\href@noop {}
  {\bibfield  {journal} {\bibinfo  {journal} {AIP Advances}\ }\textbf {\bibinfo
  {volume} {10}},\ \bibinfo {pages} {045315} (\bibinfo {year}
  {2020})}\BibitemShut {NoStop}%
\bibitem [{\citenamefont {Zhang}\ \emph {et~al.}(2021)\citenamefont {Zhang},
  \citenamefont {Shinokita}, \citenamefont {Watanabe}, \citenamefont
  {Taniguchi}, \citenamefont {Miyauchi},\ and\ \citenamefont
  {Matsuda}}]{Zhang2021}%
  \BibitemOpen
  \bibfield  {author} {\bibinfo {author} {\bibfnamefont {Y.}~\bibnamefont
  {Zhang}}, \bibinfo {author} {\bibfnamefont {K.}~\bibnamefont {Shinokita}},
  \bibinfo {author} {\bibfnamefont {K.}~\bibnamefont {Watanabe}}, \bibinfo
  {author} {\bibfnamefont {T.}~\bibnamefont {Taniguchi}}, \bibinfo {author}
  {\bibfnamefont {Y.}~\bibnamefont {Miyauchi}},\ and\ \bibinfo {author}
  {\bibfnamefont {K.}~\bibnamefont {Matsuda}},\ }\bibfield  {title} {\bibinfo
  {title} {Magnetic field induced inter-valley trion dynamics in monolayer 2{D}
  semiconductor},\ }\href@noop {} {\bibfield  {journal} {\bibinfo  {journal}
  {Adv. Funct. Mater.}\ }\textbf {\bibinfo {volume} {31}},\ \bibinfo {pages}
  {2006064} (\bibinfo {year} {2021})}\BibitemShut {NoStop}%
\bibitem [{\citenamefont {Seyler}\ \emph {et~al.}(2018)\citenamefont {Seyler},
  \citenamefont {Zhong}, \citenamefont {Huang}, \citenamefont {Linpeng},
  \citenamefont {Wilson}, \citenamefont {Taniguchi}, \citenamefont {Watanabe},
  \citenamefont {Yao}, \citenamefont {Xiao}, \citenamefont {McGuire},
  \citenamefont {Fu},\ and\ \citenamefont {Xu}}]{Seyler2018}%
  \BibitemOpen
  \bibfield  {author} {\bibinfo {author} {\bibfnamefont {K.~L.}\ \bibnamefont
  {Seyler}}, \bibinfo {author} {\bibfnamefont {D.}~\bibnamefont {Zhong}},
  \bibinfo {author} {\bibfnamefont {B.}~\bibnamefont {Huang}}, \bibinfo
  {author} {\bibfnamefont {X.}~\bibnamefont {Linpeng}}, \bibinfo {author}
  {\bibfnamefont {N.~P.}\ \bibnamefont {Wilson}}, \bibinfo {author}
  {\bibfnamefont {T.}~\bibnamefont {Taniguchi}}, \bibinfo {author}
  {\bibfnamefont {K.}~\bibnamefont {Watanabe}}, \bibinfo {author}
  {\bibfnamefont {W.}~\bibnamefont {Yao}}, \bibinfo {author} {\bibfnamefont
  {D.}~\bibnamefont {Xiao}}, \bibinfo {author} {\bibfnamefont {M.~A.}\
  \bibnamefont {McGuire}}, \bibinfo {author} {\bibfnamefont {K.~M.~C.}\
  \bibnamefont {Fu}},\ and\ \bibinfo {author} {\bibfnamefont {X.}~\bibnamefont
  {Xu}},\ }\bibfield  {title} {\bibinfo {title} {Valley manipulation by
  optically tuning the magnetic proximity effect in {WS}e$_2$/{C}r{I}$_3$
  heterostructures},\ }\href@noop {} {\bibfield  {journal} {\bibinfo  {journal}
  {Nano Letters}\ }\textbf {\bibinfo {volume} {18}},\ \bibinfo {pages} {3823}
  (\bibinfo {year} {2018})}\BibitemShut {NoStop}%
\bibitem [{\citenamefont {Benítez}\ \emph {et~al.}(2018)\citenamefont
  {Benítez}, \citenamefont {Sierra}, \citenamefont {Torres}, \citenamefont
  {Arrighi}, \citenamefont {Bonell}, \citenamefont {Costache},\ and\
  \citenamefont {Valenzuela}}]{Benitez2018}%
  \BibitemOpen
  \bibfield  {author} {\bibinfo {author} {\bibfnamefont {L.~A.}\ \bibnamefont
  {Benítez}}, \bibinfo {author} {\bibfnamefont {J.~F.}\ \bibnamefont
  {Sierra}}, \bibinfo {author} {\bibfnamefont {W.~S.}\ \bibnamefont {Torres}},
  \bibinfo {author} {\bibfnamefont {A.}~\bibnamefont {Arrighi}}, \bibinfo
  {author} {\bibfnamefont {F.}~\bibnamefont {Bonell}}, \bibinfo {author}
  {\bibfnamefont {M.~V.}\ \bibnamefont {Costache}},\ and\ \bibinfo {author}
  {\bibfnamefont {S.~O.}\ \bibnamefont {Valenzuela}},\ }\bibfield  {title}
  {\bibinfo {title} {Strongly anisotropic spin relaxation in
  graphene-transition metal dichalcogenide heterostructures at room
  temperature},\ }\href@noop {} {\bibfield  {journal} {\bibinfo  {journal}
  {Nat. Phys.}\ }\textbf {\bibinfo {volume} {14}},\ \bibinfo {pages} {303}
  (\bibinfo {year} {2018})}\BibitemShut {NoStop}%
\bibitem [{\citenamefont {Sierra}\ \emph {et~al.}(2021)\citenamefont {Sierra},
  \citenamefont {Fabian}, \citenamefont {Kawakami}, \citenamefont {Roche},\
  and\ \citenamefont {Valenzuela}}]{Sierra2021}%
  \BibitemOpen
  \bibfield  {author} {\bibinfo {author} {\bibfnamefont {J.~F.}\ \bibnamefont
  {Sierra}}, \bibinfo {author} {\bibfnamefont {J.}~\bibnamefont {Fabian}},
  \bibinfo {author} {\bibfnamefont {R.~K.}\ \bibnamefont {Kawakami}}, \bibinfo
  {author} {\bibfnamefont {S.}~\bibnamefont {Roche}},\ and\ \bibinfo {author}
  {\bibfnamefont {S.~O.}\ \bibnamefont {Valenzuela}},\ }\bibfield  {title}
  {\bibinfo {title} {Van der {W}aals heterostructures for spintronics and
  opto-spintronics},\ }\href@noop {} {\bibfield  {journal} {\bibinfo  {journal}
  {Nat. Nanotech.}\ }\textbf {\bibinfo {volume} {16}},\ \bibinfo {pages} {856}
  (\bibinfo {year} {2021})}\BibitemShut {NoStop}%
\bibitem [{\citenamefont {Dal~Conte}\ \emph {et~al.}(2015)\citenamefont
  {Dal~Conte}, \citenamefont {Bottegoni}, \citenamefont {Pogna}, \citenamefont
  {De~Fazio}, \citenamefont {Ambrogio}, \citenamefont {Bargigia}, \citenamefont
  {D'Andrea}, \citenamefont {Lombardo}, \citenamefont {Bruna}, \citenamefont
  {Ciccacci}, \citenamefont {Ferrari}, \citenamefont {Cerullo},\ and\
  \citenamefont {Finazzi}}]{DalConte2015}%
  \BibitemOpen
  \bibfield  {author} {\bibinfo {author} {\bibfnamefont {S.}~\bibnamefont
  {Dal~Conte}}, \bibinfo {author} {\bibfnamefont {F.}~\bibnamefont
  {Bottegoni}}, \bibinfo {author} {\bibfnamefont {E.~A.~A.}\ \bibnamefont
  {Pogna}}, \bibinfo {author} {\bibfnamefont {D.}~\bibnamefont {De~Fazio}},
  \bibinfo {author} {\bibfnamefont {S.}~\bibnamefont {Ambrogio}}, \bibinfo
  {author} {\bibfnamefont {I.}~\bibnamefont {Bargigia}}, \bibinfo {author}
  {\bibfnamefont {C.}~\bibnamefont {D'Andrea}}, \bibinfo {author}
  {\bibfnamefont {A.}~\bibnamefont {Lombardo}}, \bibinfo {author}
  {\bibfnamefont {M.}~\bibnamefont {Bruna}}, \bibinfo {author} {\bibfnamefont
  {F.}~\bibnamefont {Ciccacci}}, \bibinfo {author} {\bibfnamefont {A.~C.}\
  \bibnamefont {Ferrari}}, \bibinfo {author} {\bibfnamefont {G.}~\bibnamefont
  {Cerullo}},\ and\ \bibinfo {author} {\bibfnamefont {M.}~\bibnamefont
  {Finazzi}},\ }\bibfield  {title} {\bibinfo {title} {Ultrafast valley
  relaxation dynamics in monolayer {M}o{S}$_2$ probed by nonequilibrium optical
  techniques},\ }\href@noop {} {\bibfield  {journal} {\bibinfo  {journal}
  {Phys. Rev. B}\ }\textbf {\bibinfo {volume} {92}},\ \bibinfo {pages} {235425}
  (\bibinfo {year} {2015})}\BibitemShut {NoStop}%
\bibitem [{\citenamefont {Gilardoni}\ \emph {et~al.}(2021)\citenamefont
  {Gilardoni}, \citenamefont {Hendriks}, \citenamefont {van~der Wal},\ and\
  \citenamefont {Guimar\~aes}}]{Gilardoni2021}%
  \BibitemOpen
  \bibfield  {author} {\bibinfo {author} {\bibfnamefont {C.~M.}\ \bibnamefont
  {Gilardoni}}, \bibinfo {author} {\bibfnamefont {F.}~\bibnamefont {Hendriks}},
  \bibinfo {author} {\bibfnamefont {C.~H.}\ \bibnamefont {van~der Wal}},\ and\
  \bibinfo {author} {\bibfnamefont {M.~H.~D.}\ \bibnamefont {Guimar\~aes}},\
  }\bibfield  {title} {\bibinfo {title} {Symmetry and control of
  spin-scattering processes in two-dimensional transition metal
  dichalcogenides},\ }\href@noop {} {\bibfield  {journal} {\bibinfo  {journal}
  {Phys. Rev. B}\ }\textbf {\bibinfo {volume} {103}},\ \bibinfo {pages}
  {115410} (\bibinfo {year} {2021})}\BibitemShut {NoStop}%
\bibitem [{\citenamefont {Schwemmer}\ \emph {et~al.}(2017)\citenamefont
  {Schwemmer}, \citenamefont {Nagler}, \citenamefont {Hanninger}, \citenamefont
  {Schüller},\ and\ \citenamefont {Korn}}]{Schwemmer2017}%
  \BibitemOpen
  \bibfield  {author} {\bibinfo {author} {\bibfnamefont {M.}~\bibnamefont
  {Schwemmer}}, \bibinfo {author} {\bibfnamefont {P.}~\bibnamefont {Nagler}},
  \bibinfo {author} {\bibfnamefont {A.}~\bibnamefont {Hanninger}}, \bibinfo
  {author} {\bibfnamefont {C.}~\bibnamefont {Schüller}},\ and\ \bibinfo
  {author} {\bibfnamefont {T.}~\bibnamefont {Korn}},\ }\bibfield  {title}
  {\bibinfo {title} {Long-lived spin polarization in n-doped {M}o{S}e$_2$
  monolayers},\ }\href@noop {} {\bibfield  {journal} {\bibinfo  {journal}
  {Appl. Phys. Lett.}\ }\textbf {\bibinfo {volume} {111}},\ \bibinfo {pages}
  {082404} (\bibinfo {year} {2017})}\BibitemShut {NoStop}%
\bibitem [{\citenamefont {Mai}\ \emph {et~al.}(2014)\citenamefont {Mai},
  \citenamefont {Barrette}, \citenamefont {Yu}, \citenamefont {Semenov},
  \citenamefont {Kim}, \citenamefont {Cao},\ and\ \citenamefont
  {Gundogdu}}]{Mai2014}%
  \BibitemOpen
  \bibfield  {author} {\bibinfo {author} {\bibfnamefont {C.}~\bibnamefont
  {Mai}}, \bibinfo {author} {\bibfnamefont {A.}~\bibnamefont {Barrette}},
  \bibinfo {author} {\bibfnamefont {Y.}~\bibnamefont {Yu}}, \bibinfo {author}
  {\bibfnamefont {Y.~G.}\ \bibnamefont {Semenov}}, \bibinfo {author}
  {\bibfnamefont {K.~W.}\ \bibnamefont {Kim}}, \bibinfo {author} {\bibfnamefont
  {L.}~\bibnamefont {Cao}},\ and\ \bibinfo {author} {\bibfnamefont
  {K.}~\bibnamefont {Gundogdu}},\ }\bibfield  {title} {\bibinfo {title}
  {Many-body effects in valleytronics: Direct measurement of valley lifetimes
  in single-layer {M}o{S}$_2$},\ }\href@noop {} {\bibfield  {journal} {\bibinfo
   {journal} {Nano Lett.}\ }\textbf {\bibinfo {volume} {14}},\ \bibinfo {pages}
  {202} (\bibinfo {year} {2014})}\BibitemShut {NoStop}%
\bibitem [{\citenamefont {Schellekens}(2014)}]{Schellekens2014}%
  \BibitemOpen
  \bibfield  {author} {\bibinfo {author} {\bibfnamefont {A.~J.}\ \bibnamefont
  {Schellekens}},\ }\emph {\bibinfo {title} {Manipulating spins : novel methods
  for controlling magnetization dynamics on the ultimate timescale}},\ \href
  {https://doi.org/10.6100/IR770953} {Ph.D. thesis},\ \bibinfo  {school}
  {Eindhoven University of Technology} (\bibinfo {year} {2014}),\ \bibinfo
  {note} {{I}SBN: 978-90-386-3586-6}\BibitemShut {NoStop}%
\end{thebibliography}%


\onecolumngrid
\clearpage

\section*{Supplementary Information: \\ - \\ $\large \text{Magnetic field control of light-induced spin accumulation in monolayer MoSe}_2$}
\renewcommand\thefigure{S\arabic{figure}}
\renewcommand{\figurename}{Figure}

\section{Details on the experimental setup}

\begin{figure}[ht!]
    \centering
    \includegraphics[scale=0.9]{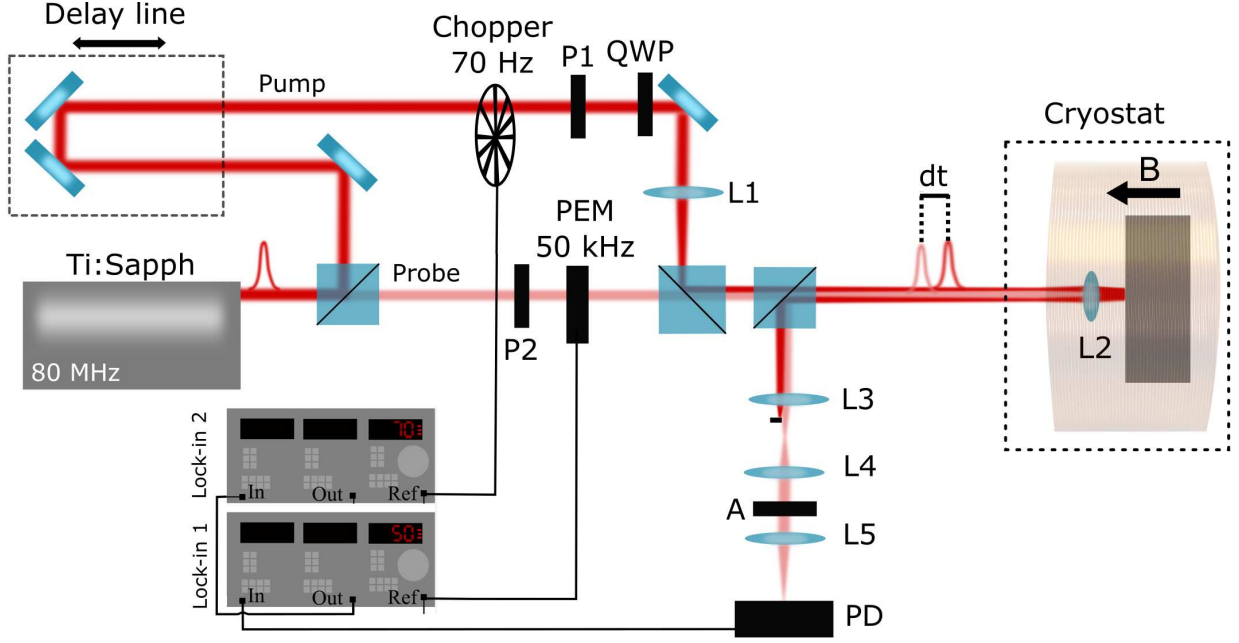}
    \caption{Experimental setup used for our time-resolved measurements.}
    \label{FigSI1:Setup}
\end{figure}

For measuring time-resolved data we used a single-color (degenerated) pump-probe technique in a double modulation configuration.
Figure \ref{FigSI1:Setup} pictures our experimental setup with the key elements represented on it. 
For excitation, we used a tunable Ti:sapphire laser (Mai Tai - Spectra Physics) with a repetition rate of 80 MHz and a pulse width $<$ 300 fs. 
With a beam-splitter and a set of neutral density filters (not shown in Figure \ref{FigSI1:Setup}), we set the pump and probe beams to a 4:1 fluence ratio with a $\sim$ 6 $\mu$J/cm$^2$ probe fluence. 
The pump is set to circularly polarized light ($\sigma _\pm$) with a linear polarizer (P1) and a quarter wave plate (QWP) after passing through a chopper with a frequency of 70 Hz.
For blocking the pump beam after the reflection on the sample we use the set of lenses L1, L3, and L4 that help focus the beam in a different place between L3 and L4, on its way back from the cryostat.
The lens L1 also ensures a larger spot size for the pump beam at the sample, due to a different focal point.
The probe beam is linearly polarized (P2) and its polarization is modulated by a photoelastic modulator (PEM) at a frequency $\Omega\ =$ 50 kHz. 
The beam is detected on the way back from the sample by an amplified photodetector (PD) after passing through a linear polarizer (A), set cross-polarized with the probe for detecting the Kerr rotation signal.

The presence of the chopper in the pump path and the PEM in the probe path has the purpose of improving the signal-to-noise ratio, also known as the double modulation technique.
The PEM is set with its fast axis at 45$\degree$ to the axis of polarization of the incident beam that makes the polarization of the probe oscillate between circularly and linearly polarized with a frequency $2\Omega$. 
The difference in the modulation frequencies allows us to separate the signal coming from each beam.
With this configuration, the reflected signal is sent to one lock-in amplifier with the reference to the PEM oscillation frequency, which helps clean the signal coming from the probe beam.
The output of the filtered signal is sent to a second lock-in amplifier that has its reference on the chopper frequency and helps to improve the signal from the probe which is affected by the pump.  
Using the Jones matrix formalism, it is possible to determine that the intensity of the signal at the second harmonic of the PEM oscillation frequency is related to the Kerr rotation as $I_{2\Omega}=\theta _K RJ_2(\rho)\cos (2\Omega t)$ \cite{Schellekens2014, Guimaraes2018}, where $\theta _K$ the Kerr rotation, $R$ is the average reflected intensity, $J_2$ the spherical Bessel function of the first kind and $\rho $ = 3.041 is the retardation given by the PEM to optimize the signal. 
Therefore, the first lock-in amplifier is set to detect and amplify the second harmonic which is proportional to the Kerr rotation signal.

\section{Photoluminescence at high magnetic fields}

\begin{figure}[!htb]
    \centering
    \includegraphics[width=\linewidth]{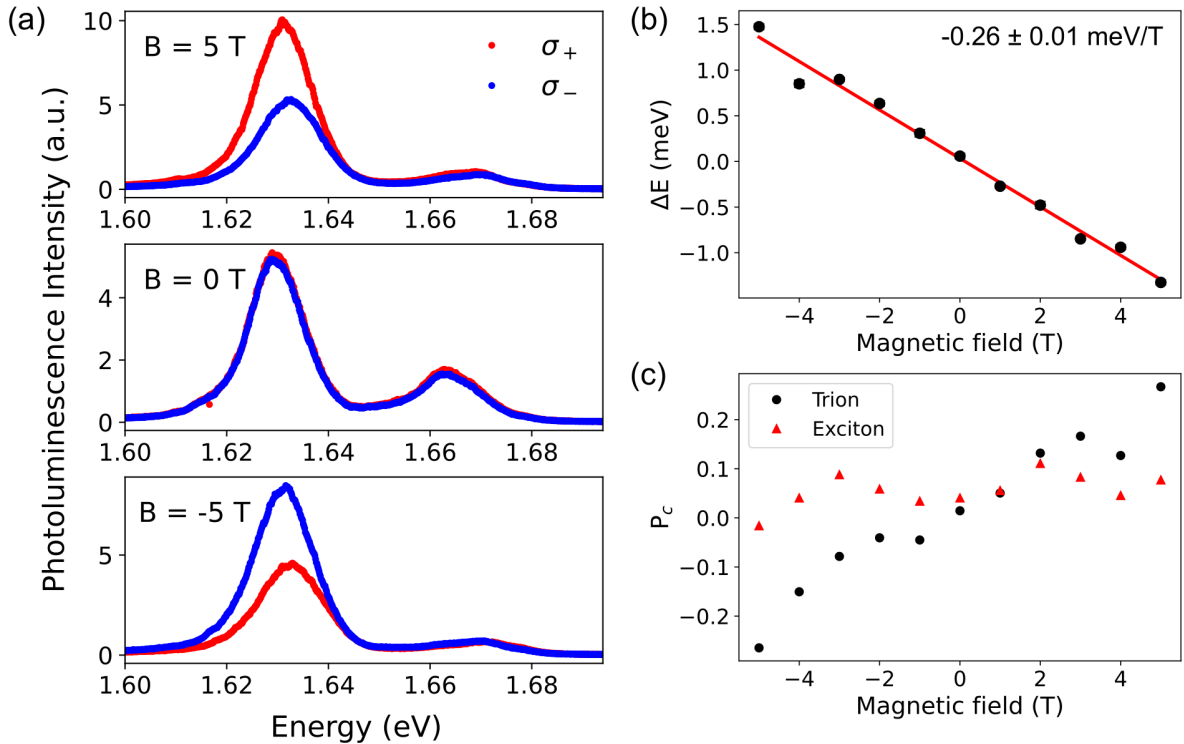}
    \caption{Photoluminescence spectra of 1L MoSe$_2$ at different magnetic fields for the two polarization excitations. (b) Valley-Zeeman splitting for the trion peak, calculated as the difference of the energies $E_{\sigma_+}-E_{\sigma_-}$. (c) Degree of circular polarization (Pc) of excitons and trions at different magnetic fields.}
    \label{FigSI1:PL}
\end{figure}

Photoluminescence measurements were performed with $\lambda =$ 690 nm laser diode with a power of 100 $\mu$W at a temperature of 6 K.
For controlling the polarization, we set the fast axis of a quarter wave plate (QWP) aligned to the polarization of the excitation light.
When the emitted light from the sample, $E_+\hat{\sigma}_++E_-\hat{\sigma}_-$, with $\hat{\sigma}_+$ the Jones matrix associated to the circularly polarized light, passes again through the waveplate, the fast axis is no longer aligned with the polarization of the light.
The latter will decompose most of the signal of the circularly polarized states at $\pm \pi/4$ with respect to the axis of the QWP.
Therefore, by placing an analyzer (polarizer) at those angles, we can independently measure the intensity $\vert E_+\vert^2$ and $\vert E_-\vert^2$.

Figure \ref{FigSI1:PL}a shows the photoluminescence spectra of our sample at different magnetic fields B = 0, $\pm$5 T.
Each spectrum was measured at the two polarizations, $\sigma_+$ (red) and $\sigma_-$ (blue).
The two characteristic peaks of MoSe$_2$ are observed: one coming from exciton recombination (1.664 eV) and the other from the emission of trions (1.630 eV).
For analyzing the valley-Zeeman shifting we fitted the trion peak for both polarizations and plotted the difference of the peak centers $E_{\sigma_+}$-$E_{\sigma_-}$ as a function of the magnetic field (Figure \ref{FigSI1:PL}b).
We obtained a linear behavior with a slope of ($-0.26 \pm 0.01$) meV/T, indicating a $g$-factor of -4.5, which is in good agreement with previous reports in the literature \cite{Macneill2015,Koperski2019,Wozniak2020}.
Figure \ref{FigSI1:PL}c shows the degree of circular polarization of the exciton and trion peaks as a function of the magnetic field calculated as the difference of the integrated intensity at the two different polarizations $P_c=(I_{\sigma_+}-I_{\sigma_-})/(I_{\sigma_+}+I_{\sigma_-})$.

\clearpage

\section{Fast decay process in TRKR vs. Magnetic field}

\begin{figure}[h!]
    \centering
    \includegraphics[width=\linewidth]{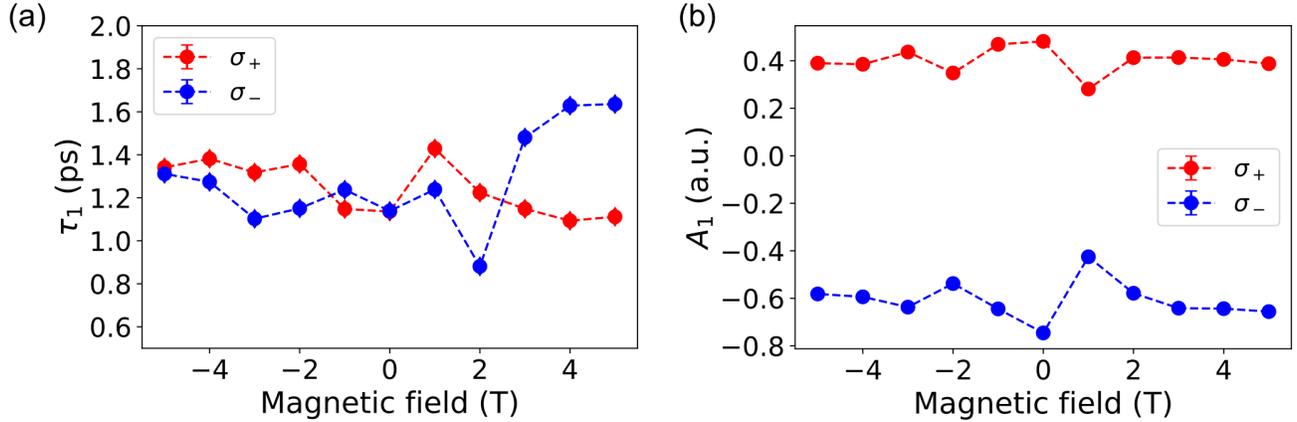}
    \caption{Extracted fast decay times $\tau _1$ and (b) amplitudes A$_1$ from the fits of TRKR presented in the main manuscript.}
    \label{FigSI4:Fast_2exp}
\end{figure}

Figure \ref{FigSI4:Fast_2exp} shows the fast decay times $\tau_1$ and corresponding amplitudes $A_1$ obtained from the biexponential decay fits of the Kerr rotation angles measured at different magnetic fields.

\section{Spin dynamics: wavelength dependence}

\begin{figure}[ht!]
    \centering
    \includegraphics[scale=0.9]{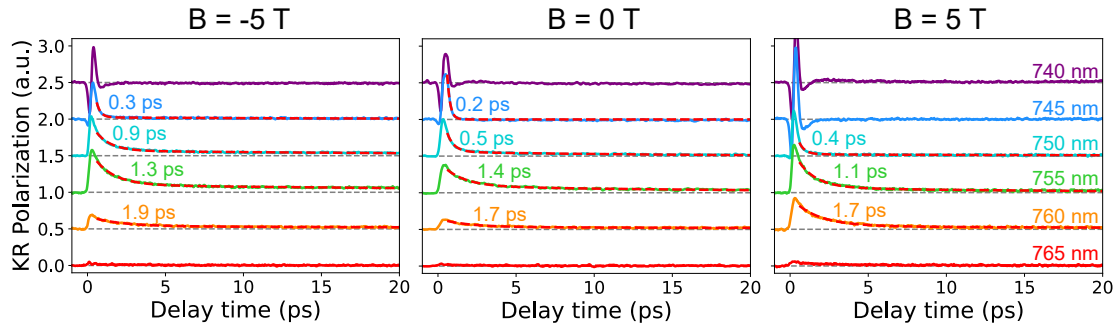}
    \caption{Kerr rotation polarization of our sample at different wavelengths studied with an out-of-plane magnetic field B=$\pm$5 T and at B = 0 T. Extracted lifetimes are shown at each plot.}
    \label{FigSI2:KR_Pol_wl}
\end{figure}

For keeping track of the wavelength dependence of the Kerr rotation with the magnetic field, we also measured the TRKR at the two polarizations ($\sigma_\pm$) for different wavelengths at B = 0 T, B = $\pm$5 T. 
To reduce spurious effects and simplify the analysis, we calculate the Kerr rotation polarization defined as ($\theta_K (\sigma_+)-\theta_K(\sigma_-)$)/2.
The results are shown in Figure \ref{FigSI2:KR_Pol_wl}.
In all the cases, we observe that the stronger and long-lived signals are associated with the wavelengths between $\lambda$ = 755 nm and 760 nm.
Therefore, the magnetic field does not induce clear changes in the KR signal close to exciton resonances nor change (significantly) the resonance close to the trion energy peak. 

\clearpage

\section{Rate equation Model -Different parameters}

\begin{figure}[h!]
    \centering
    \includegraphics[scale=0.95]{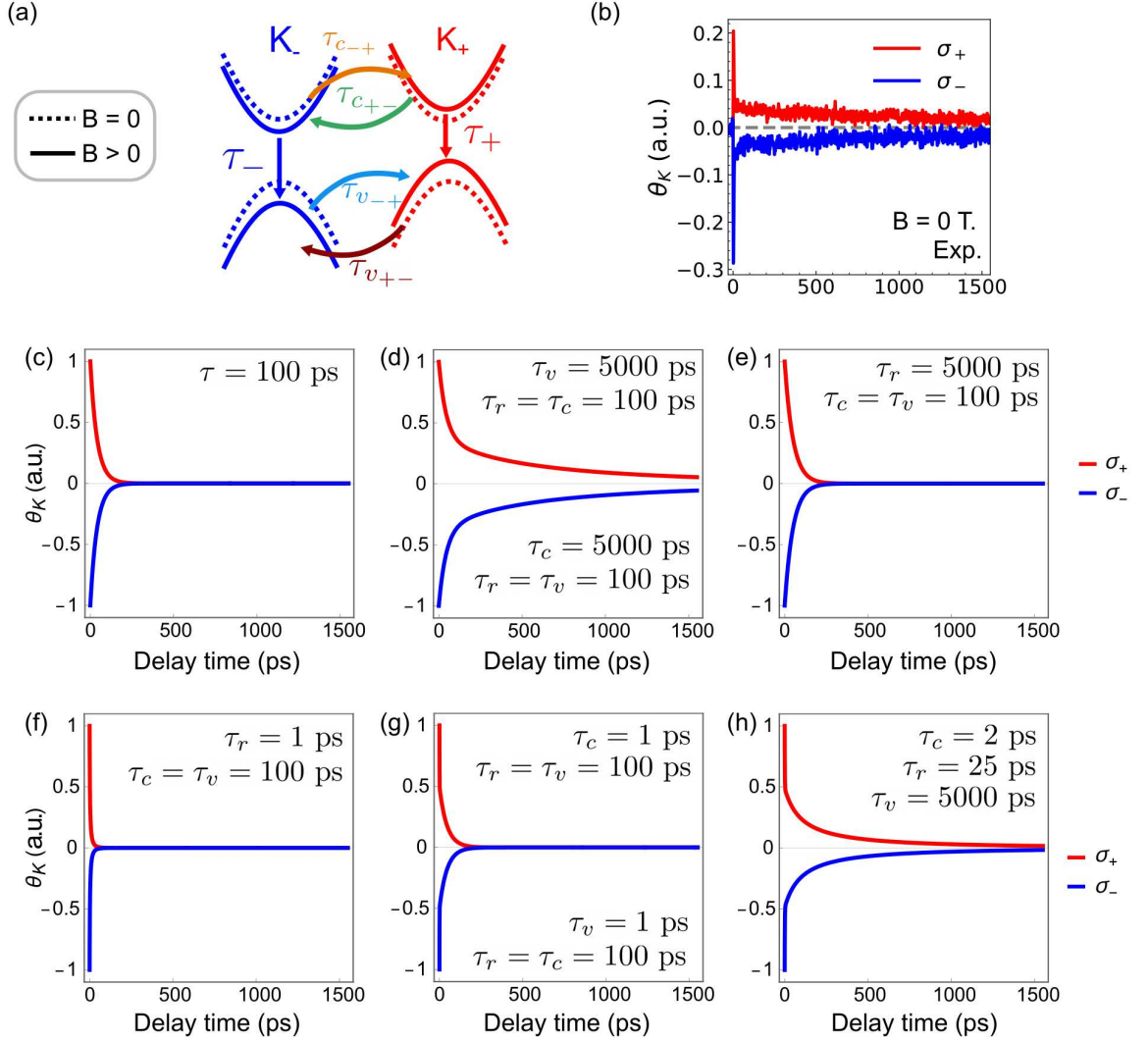}
    \caption{(a) Representation of the valley-Zemman shifting and carrier transfer process of the bands at the vallleys $K_\pm$ under a positive magnetic field. (b) Experimental result of TRKR in MoSe$_2$ at zero magnetic field for the two pump polarizations $\sigma_\pm$ at T = 6 K. (c-h) Simulation of the TRKR using the model presented in the main text. The parameters for the calculation are presented in each figure. When two sets of parameters are presented on the same figure, it indicates that the same result is obtained for both cases.}
    \label{FigSI7:ModelB0}
\end{figure}

Figure \ref{FigSI7:ModelB0}a shows all the different decay paths considered in our model.
When a magnetic field is applied, valley degeneracy breaking can affect the recombination and scattering times of excited spins at each valley.
To start exploring the effect of such asymmetry, we first pursued the reproduction of our results at zero magnetic field, Figure \ref{FigSI7:ModelB0}b.
In Figure \ref{FigSI7:ModelB0}(c-h) we present the TRKR calculated for different sets of parameters.
As for zero magnetic field valley degeneracy is still present, equal scattering and recombination rates are considered for both valleys: $\tau_+=\tau_-=\tau_r$, $\tau_{v_{+-}}=\tau_{v_{-+}}=\tau_{v}$, and $\tau_{c_{+-}}=\tau_{c_{-+}}=\tau_{c}$. 

\begin{figure}[h!]
    \centering
    \includegraphics[scale=1]{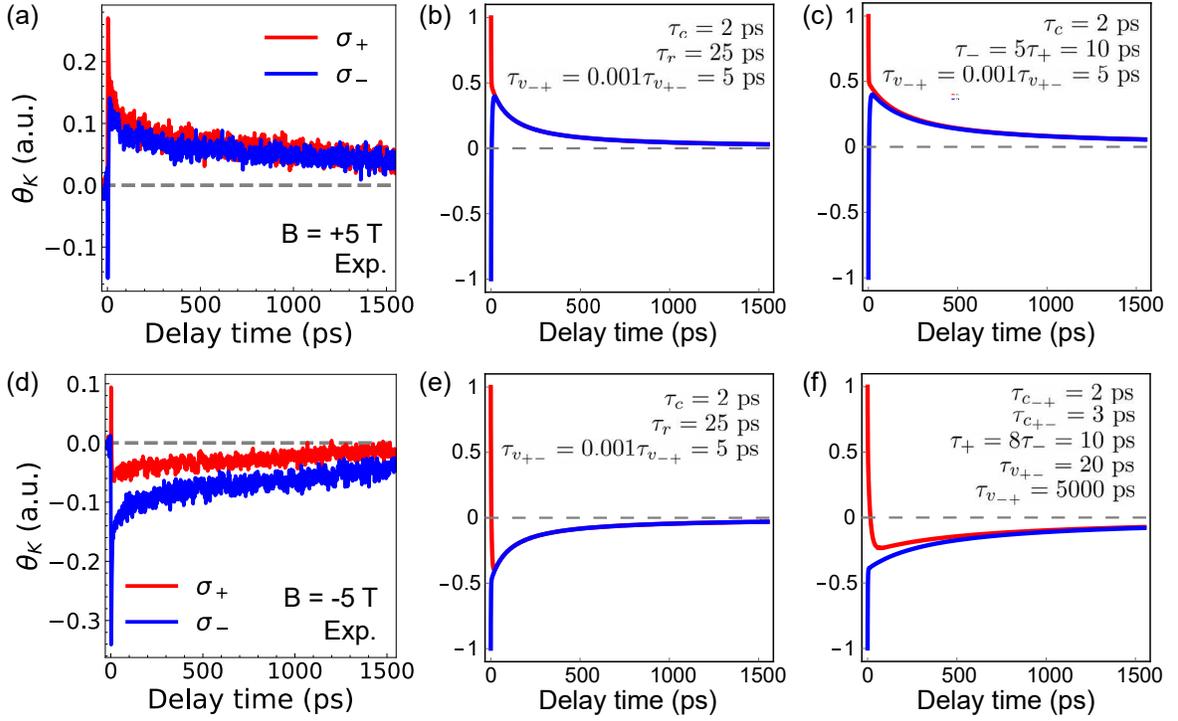}
    \caption{(a) Experimental TRKR of MoSe$_2$ at B = 5 T. (b-c) Calculated TRKR for modeling (a). (d) Experimental result at B = -5 T. (e-f) Simulated TRKR for modeling the results shown in (d)}
    \label{FigSI7:ModelB5}
\end{figure}

With the obtained parameters at zero magnetic field (Figure \ref{FigSI7:ModelB0}h), we proceed to study the effect of the magnetic field introducing the assymetry in the decay times.
Figure \ref{FigSI7:ModelB5} show the results of TRKR calculted with different parameters for reproducing our experimental results at B = $\pm$ 5 T.

\newpage

\section{TRKR vs. B - Different set of measurements}

\begin{figure}[h!]
    \centering
    \includegraphics[scale=1]{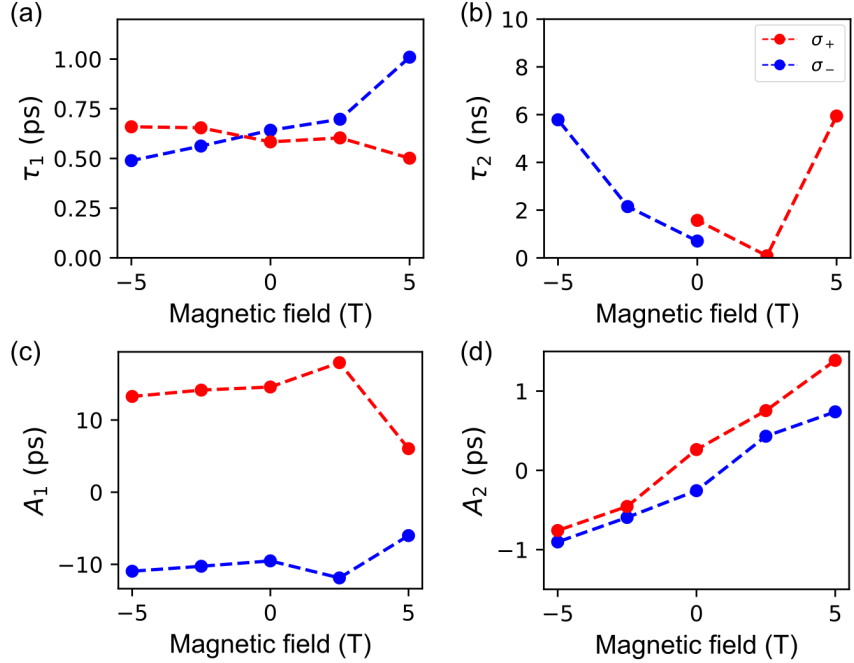}
    \caption{(a-b) TRKR decay times and (c-d) amplitudes extracted at a different point of the sample, in a different cool down of the same sample, with a different alignment with $\lambda$ = 755 nm, $F_{pump}$ = 100 $\mu J/cm^2$, $F_{probe}$ = 10 $\mu J/cm^2$, T = 6 K.}
    \label{FigSI8:Old_fits}
\end{figure}
\newpage


\section{Further discussion on fitting the decay process}

\begin{figure}[h!]
    \centering
    \includegraphics[scale=1]{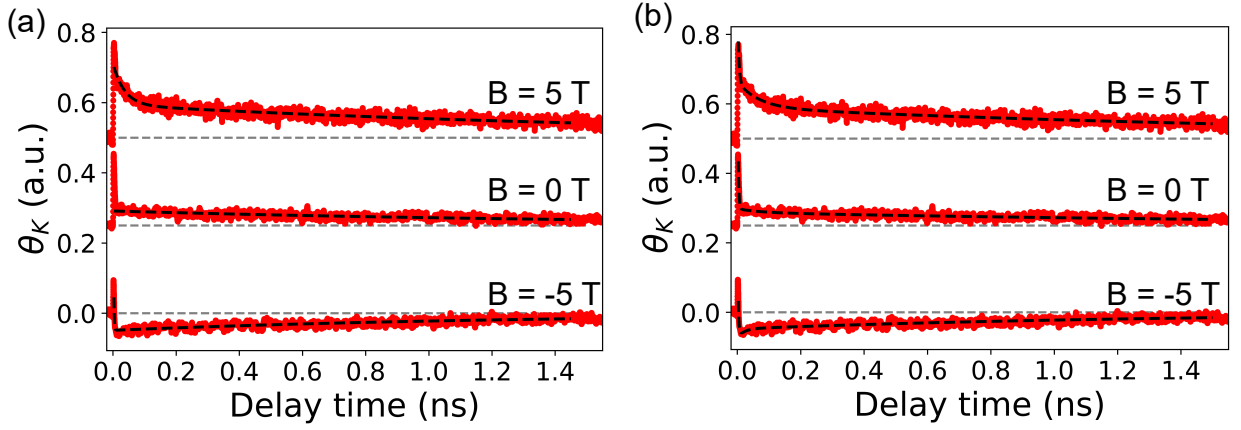}
    \caption{(a) TRKR at different magnetic fields for a $\sigma_+$ excitation fitted with a bi-exponential decay and (b) a three-exponential decay (black dashed line).}
    \label{FigSI5:TRKR_2vs3_exp_decay}
\end{figure}

For fitting the spin dynamics in the main document we considered a simple approach of two exponential decay, as it is within the scope of our paper.
For the fits, we use the data measured in two ranges with different acquisition parameters.
One for short-time scale analysis up to 20 ps, acquired at 0.1 ps/s, and a long one for the slow decay process up to 1550 ps, with a velocity of acquisition of 3 ps/s.
For each data set is possible to accurately fit either the fast or the slow decay process. 
Nevertheless, fits for intermediate decay times (tens of ps) can fail for some curves (see Figure \ref{FigSI5:TRKR_2vs3_exp_decay}a).
Here, we present an alternative three-decay process fit  ($A_1e^{-t/\tau_1}+A_2e^{-t/\tau_2}+A_3e^{-t/\tau_3}$) over the long-range data set that helps to elucidate the dynamics of the intermediate process and its effect on the results of the two-exponential decay.

\begin{figure}[h!]
    \centering
    \includegraphics[width=\linewidth]{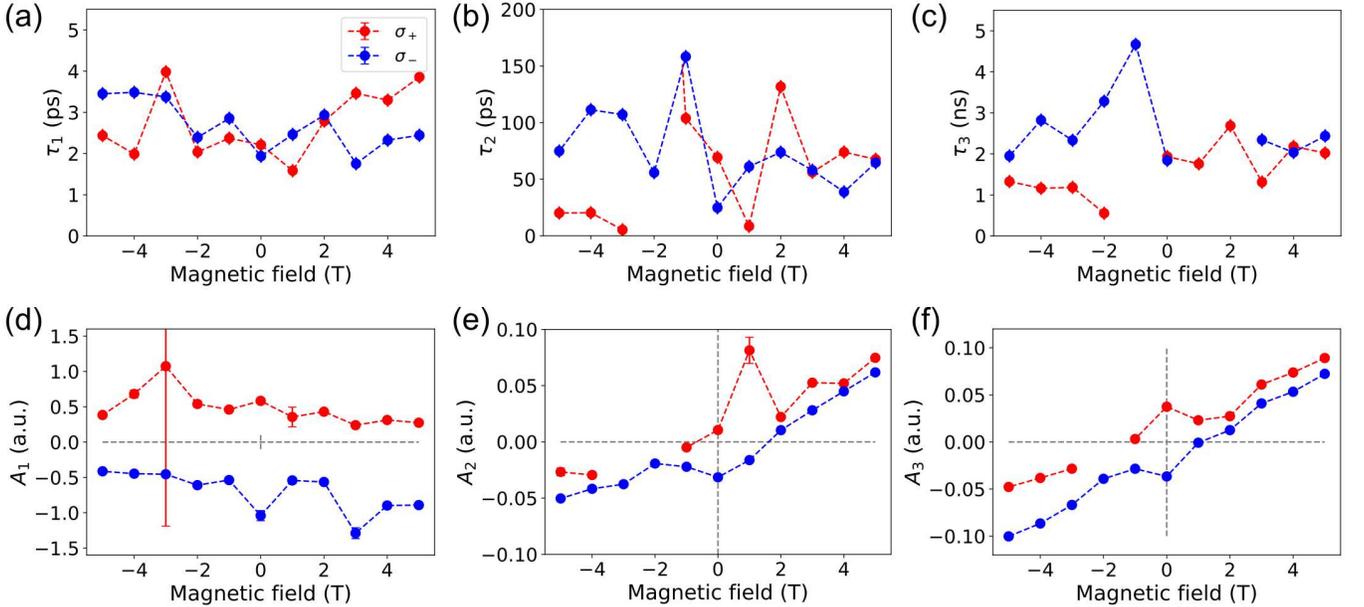}
    \caption{Fitting parameters extracted from a three exponential decay fit.}
    \label{FigSI6:3decay_param}
\end{figure}

Figure \ref{FigSI5:TRKR_2vs3_exp_decay}b presents the fits of the TRKR when exciting with polarization $\sigma_+$ at three different magnetic fields. 
In Figure \ref{FigSI6:3decay_param} we present the extracted lifetimes and amplitudes of each decay process.
We observe a consistent result for the fast and slow decay processes when compared with the data presented with the two-exponential decay case.
For $\tau_2$ we observed lifetimes of tens to a hundred picoseconds, which is consistent with the trion recombination lifetime reported in the literature \cite{Schwemmer2017, Godde2016, Hsu2015, Wang2015}. 
Also, for the amplitude $A_2$, we observe a linear behavior with the magnetic field as observed for $A_3$.

\section{Time-Resolved Differential Reflectivity at B = 0 T}

\begin{figure}[h!]
    \centering
    \includegraphics[scale=0.8]{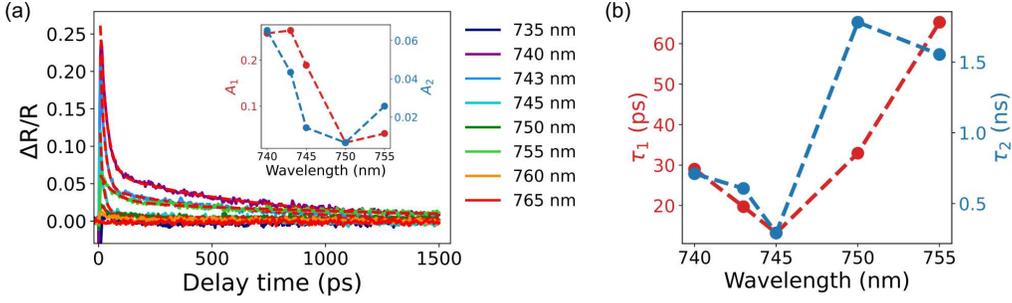}
    \caption{(a) Time-resolved differential reflectivity studied at different wavelengths and their fits (red dashed lines). Inset: extracted amplitudes $A_i$ and; (b) extracted lifetimes $\tau _i$ of the bi-exponential fit.}
    \label{FigSI3:TRDR_wl}
\end{figure}

\end{document}